\documentclass[preprint,times,sort&compress]{elsarticle}

\pdfoutput=1

\journal{ArXiv}

\usepackage{url}
\usepackage[breaklinks,hidelinks]{hyperref}

\usepackage{graphicx}

\usepackage{algorithmic}
\usepackage{algorithm}

\usepackage{amsmath}
\usepackage{mathtools}

\usepackage{multirow}
\usepackage{booktabs}
\usepackage{makecell}

\usepackage[format=hang,labelformat=parens,subrefformat=parens]{subcaption}
\usepackage[binary-units]{siunitx}

\usepackage[T1]{fontenc}

\usepackage{pdflscape}

\usepackage[acronym]{glossaries}
\newacronym{SM}{SM}{Standard Model}
\newacronym{TOE}{TOE}{Theory Of Everything}
\newacronym{LHC}{LHC}{Large Hadron Collider}
\newacronym{CERN}{CERN}{the European Organization for Nuclear Research}
\newacronym{PM}{PM}{Post Mortem}
\newacronym{PMS}{PM System}{\glsentrylong{PM} System}
\newacronym{PMA}{PMA}{\glsentrylong{PM} Analysis}
\newacronym{PMAF}{PMA Framework}{\glsentrylong{PMA} Framework}
\newacronym{PMF}{PM Framework}{\glsentrylong{PM} Framework}
\newacronym{LEP}{LEP}{Large Electron-Positron Collider}
\newacronym{PS}{PS}{Proton Synchrotron}
\newacronym{SPS}{SPS}{Super Proton Synchrotron}
\newacronym{ATLAS}{ATLAS}{A Toroidal LHC ApparatuS}
\newacronym{TRT}{TRT}{Transition Radiation Tracker}
\newacronym{SCT}{SCT}{SemiConductor Tracker}
\newacronym{CMS}{CMS}{Compact Muon Solenoid}
\newacronym{ALICE}{ALICE}{A Large Ion Collider Experiment} 
\newacronym{LHCb}{LHCb}{Large Hadron Collider beauty experiment}
\newacronym{LCG}{LCG}{\gls{LHC} Computing Grid}
\newacronym{SUSY}{SUSY}{SUperSYmmetry}
\newacronym{SSB}{SSB}{Spontaneous Symmetry Breaking}
\newacronym{QPS}{QPS}{Quench Protection System}
\newacronym{MPS}{MPS}{Machine Protection System}
\newacronym{BIS}{BIS}{Beam Interlock System}
\newacronym{PIS}{PIS}{Power Interlock System}
\newacronym{SDDS}{SDDS}{Self-Describing Data Set}
\newacronym{MTF}{MTF}{Magnet Test Folder}
\newacronym{ELQA}{ELQA}{ELectrical Quality Assurance}
\newacronym{UPS}{UPS}{Uninterruptible Power Supplies}
\newacronym{BLM}{BLM}{Beam Loss Monitor}
\newacronym{BPM}{BPM}{Beam Position Monitor}
\newacronym{FMCM}{FMCM}{Fast Magnet Current change Monitor}
\newacronym{WIC}{WIC}{Warm magnets Interlock Controller}
\newacronym{AUG}{AUG}{emergency stop of electrical supplies}
\newacronym{LBDS}{LBDS}{LHC Beam Dumping System}
\newacronym{PC}{PC}{Power Converters}
\newacronym{WorldFIP}{WorldFIP}{World Factory Instrumentation Protocol}
\newacronym{LS1}{LS1}{First Long Shutdown}
\newacronym{API}{API}{Application Programming Interface}
\newacronym{REST}{REST}{Representational State Transfer}
\newacronym{JSON}{JSON}{JavaScript Object Notation}
\newacronym{CMW}{CMW}{CERN Controls Middleware}
\newacronym{RDA}{RDA}{Remote Device Access}
\newacronym{CALS}{CALS}{CERN Accelerator Logging Service}
\newacronym{RoI}{RoI}{Regions-of-Interest}

\newacronym{TE-MPE-EE}{TE-MPE-EE}{Technology Department --Machine Protection and Electrical Integrity group -- Electrical Engineering group}

\newacronym{DL}{DL}{Deep Learning}
\newacronym{LSTM}{LSTM}{Long Short-Term Memory}
\newacronym{GRU}{GRU}{Gated Recurrent Unit}
\newacronym{RMSE}{RMSE}{Root-Mean-Square Error}
\newacronym{RNN}{RNN}{Recurrent Neural Network}
\newacronym{ASIC}{ASIC}{Application-Specific Integrated Circuit}
\newacronym{FPGA}{FPGA}{Field-Programmable Gate Array}
\newacronym{FNN}{FNN}{Feed-forward Neural Network}
\newacronym{CNN}{CNN}{Convolutional Neural Network}

\begin{document}

\begin{frontmatter}

\title{Recurrent Neural Networks for anomaly detection in the Post-Mortem time series of LHC superconducting magnets}

\author[aghaddress_eit,cernaddress]{Maciej Wielgosz}
\ead{wielgosz@agh.edu.pl}

\author[aghaddress_fis]{Andrzej Skocze\'n}
\ead{skoczen@fis.agh.edu.pl}

\author[cernaddress]{Matej Mertik}
\ead{matej.mertik@cern.ch}

\address[aghaddress_eit]{Faculty of Computer Science, Electronics and Telecommunications, AGH University of Science and Technology, Krak\'ow, Poland}
\address[aghaddress_fis]{Faculty of Physics and Applied Computer Science, AGH University of Science and Technology, Krak\'ow, Poland}
\address[cernaddress]{The European Organization for Nuclear Research - CERN, CH-1211 Geneva 23 Switzerland}

\begin{abstract}
This paper presents a model based on \glsentrylong{DL} algorithms of \glsentryshort{LSTM} and \glsentryshort{GRU} for facilitating an anomaly detection in \glsentrylong{LHC} superconducting magnets. We used high resolution data available in \glsentrylong{PM} database to train a set of models and chose the best possible set of their hyper-parameters. Using \glsentrylong{DL} approach allowed to examine a vast body of data and extract the fragments which require further experts examination and are regarded as anomalies. The presented method does not require tedious manual threshold setting and operator attention at the stage of the system setup. Instead, the automatic approach is proposed, which achieves according to our experiments accuracy of \SI{99}{\percent}. This is reached for the largest dataset of 302 MB and the following architecture of the network: single layer \glsentryshort{LSTM}, 128 cells, 20 epochs of training, look\_back=16, look\_ahead=128, grid=100 and optimizer \textit{Adam}. All the experiments were run on GPU Nvidia Tesla K80.
\end{abstract}


\end{frontmatter}

%\linenumbers

\section{Introduction}
\label{section:intro}

The \gls{LHC} located at \gls{CERN} on Switzerland and France border is the largest experimental instrument which was ever built \cite{LHC_Nature}. It generates a tremendous amount of data which is later used in analysis and validation of the physics models regarding the history of the universe and the nature of the matter. Besides the data used in physics experiments, the data from the multitude of devices installed inside the \gls{LHC}, such as ones responsible for a particle beam trajectory control and stabilization of the \gls{LHC} operating parameters, is gathered. To work efficiently, those devices need to be constantly monitored and maintained and their operating parameters analyzed. As a result, each of those devices can be considered a separate system, with its own sensors and elements responsible for work control. This architecture leads to a great number of data streams depicting various systems' condition.

Some of the most vulnerable \gls{LHC} components are superconducting magnets. They are unique elements, designed and manufactured specially for the \gls{CERN}, which is why controlling their operating parameters and preventing malfunctions and failures is so important. In the \gls{CERN} history, occurrences such as \cite{cern_crash} took place, which resulted in a damage to those valuable components. As a consequence, dedicated teams, responsible for magnets maintenance and faults prevention, were formed. Members of those teams are experts in the fields of superconducting materials, cryogenic and many others and they have created models that allow to control magnets operation. Those models were hand-crafted and created from scratch and their development and adaptation is a time-consuming task, as well as requiring involvement of many people.

In this paper we attempt to automate the task of determining parameters of safe superconducting magnets' operation or at least reduce the necessary experts involvement. It should be noted that specialists cannot be removed form the process of model creation, however their work can be made easier by automating the model itself. Consequently, we try to use \glspl{RNN}, such as \gls{LSTM} and \gls{GRU}, to model electromagnets behavior.

The rest of the paper is organized as follows. Sections \ref{section:lhc} and \ref{section:rnn} provide background work about \gls{LHC} and \glspl{RNN}, respectively. System operation layer is presented in Section \ref{section:operation_layer}, with proposed method described in Section \ref{section:proposed_method}. Section \ref{section:experiments} provides the results of the experiments. Finally, the conclusions of our research are presented in Section \ref{section:conclusions}.
\section{\glsentrylong{LHC}}
\label{section:lhc}

The main objective of physics experiments carried out at the LHC is to confirm the theory known as \gls{SM}. Despite the \gls{SM} being the best description of the whole physical reality, it is 
not a \gls{TOE}. Therefore, the \gls{LHC} experiments expect to reach beyond the \gls{SM} and find a new physics. This would help to discriminate between many existing theories and to answer several tough questions about the Universe. 

In 2000, the \gls{LEP} was disasembled to make it possible to start the construction of the \gls{LHC}, the largest accelerator at \gls{CERN} \cite{LHCDesRep}. It accelerates two proton beams 
traveling in opposite directions. Therefore, the \gls{LHC} may be considered to be actually two accelerators in one assembly. Before the beams are injected into the \gls{LHC}, proton bunches are 
prepared by the \gls{PS} and \gls{SPS} accelerators, which were constructed and used at \gls{CERN} \cite{SPSdesignReport} before the \gls{LHC} was built. In each beam, there is a very large 
number of particles, which increases the probability of observing interesting collisions. A single bunch contains nominally \num{1.15d11} protons. The operation of gradual delivery of proton 
bunches to the \gls{LHC} is denoted as ``filling the machine''. It takes \num{2808} bunches altogether to fill up the \gls{LHC}. The time between bunches is \SI{25}{\nano\second}.

Upon completion of a full acceleration cycle, a velocity of the protons approximately deviates from the speed of light by about one millionth. It is hard to 
consider a value of proton velocity and therefore kinetic properties of a single proton are described by its total energy, which reaches \SI{7}{\tera\electronvolt} just before collision. The particles 
circle the \SI{27}{\kilo\metre} long beam pipe \num{11245} times per second. Particle tracks are formed by superconducting magnets working at a temperature of superfluid helium at about 
\SI{1.9}{\degree\kelvin}. Each of eight sectors of the LHC comprises about \num{154} magnets. The magnets produce a magnetic field appropriate to bend proton trajectory when they conduct an 
electrical current at the level of \SI{13}{\kilo\ampere}. In order to reach and maintain such extreme working parameters, the machine employs a dedicated helium cryogenic installation. 

The \gls{LHC} has four interaction points where bunches collide. The bunches collide every \SI{25}{\nano\second}, so there are \num{40} million of collisions every second. When two bunches collide, a number of individual proton-proton collisions occurs. Typically, there are about \num{20} events in one bunch-bunch collision. Therefore, detectors are capable of capturing 
\num{800} million of events every second. Around interaction points huge detection systems were built in order to record a complete image of all events during each collision. 
These systems are called ``detectors'' or ``experiments''. They consists of coaxial cylinders placed around the beam pipe. 
Each of the cylinders has three major layers serving different purposes. 
The innermost cylinder is a tracking system called Inner Detector. Its purpose is to record trajectories of collisions' products. 
The middle layer, called Calorimeter, measures a total energy of products. The outermost layer is a Muon Spectrometer, which allows to identify and measure the momenta of muons. 
Each of those three detector subsystems consists of several layers build with different technologies. 
The whole system is immersed in a high magnetic field parallel to the beam axis. The magnetic field bends trajectories of electrically-charged particles produced in collision. This magnetic field is
generated by a system of huge superconducting magnets installed between layers of sensors. 

The \gls{LHC} has four huge detectors: two large and versatile -- \gls{ATLAS} and \gls{CMS}, and two smaller, more specialized -- \gls{ALICE} and \gls{LHCb}.
In order to give an impression of what a particle physics experiment is, we focus on the \gls{ATLAS}, especially \gls{ATLAS} Inner Detector \cite{ATLAS2008}.

The \gls{ATLAS} detector is \SI{44}{\meter} long along the beam pipe, and the radius is \SI{11}{\meter} perpendicular to the beam pipe. The weight of the \gls{ATLAS} is about \SI{7000}{\tonne}. 
The Inner Detector of the 
\gls{ATLAS} consists of three subsystems: Pixel Detector, \gls{SCT}, and \gls{TRT}. The \gls{SCT} is a silicon microstrip tracker which comprises of \num{4088} double-sided modules. Each module
consists of \num{12} particle sensors. The sensor is a silicon die with \num{128} strips. One strip is a p-n diode with \SI{126}{\milli\meter} length. The pitch between strips is 
\SI{80}{\micro\meter}. The sensor die is bonded to a front-end electronic chip made as an \glsentryshort{ASIC}. In total there are \num{4088} $\times$ \num{12} $\times$ \num{128} = 
\num{6279168} $\approx$ \num{6,3} million of electronic channels. The modules are installed on \num{4} cylindrical barrel layers and \num{18} planar endcap discs. In total there are about 
\SI{61,1}{\meter\squared} of silicon surface. 

Those devices utilize high-end electronic solutions to be able to acquire as much as possible of each event occurring within the accelerator. Most of the detector components were customarily 
designed to meet very rigorous parameters such as very low acquisition latency and very high immunity to radiation damages. The components of the detectors which capture the signals and decide which data to 
pass on for the analysis in 
the data center \cite{nieke2015analysis} need to be very fast and are implemented in \glspl{ASIC} and \glspl{FPGA} \cite{angelucci2014FPGA}. Consequently, the design process is very tedious, 
costly and challenging. The components of the detectors which are responsible for selecting the right data are denoted as triggers and they serve a special role in distinguishing a valuable data. 
The trigger system of the \gls{ATLAS} detector is organized in three levels of fast introductory analysis. First trigger level selects about \num{100} thousand bunch crossings out of 
\num{40} million of collisions every second. The decision is undertaken during \SI{2}{\micro\second}. Already at this level a bunch crossing is divided into several \gls{RoI}, i.e. the
geographical coordinates, of those regions within the detector where its selection process has identified interesting features. Second trigger level reduces the rate to approximately \num{3.5} 
thousand events per second, with an event processing time of about \SI{40}{\milli\second}. Third trigger level is an event filter, which reduces the event rate to roughly \num{200} per second. 
This final stage is implemented using offline analysis within an event processing time of the order of \SI{4}{\second}. The output data is passed on to a data storage. 
The recorded data is investigated around the world by means of using \gls{LCG}.

\section{\glsentrylongpl{RNN}}
\label{section:rnn}

Virtually all real world phenomena may be characterized by its spacial and temporal components. The spacial ones exist in space and it is assumed that they are stationary i.e. do not develop in time. Whereas the temporal ones unfold in time and have no spacial component. This is an idealization since there are neither pure spatial nor temporal phenomena, most of them may be described as a mixture of those two different components.

There is a well-established practice in \glsentrylong{DL} applications to use \glspl{FNN} and \glspl{CNN} to address tasks dominated by a spacial component \cite{krizhevsky}. On a contrary, data which contain more temporally distributed information are usually processed by models built around \glspl{RNN}. Of course, it is possible to treat time series signals as a vector of spatial values and use \gls{FNN} or \gls{CNN} to classify them or do some regression \cite{LeCun_deep_learning_2015}.

The voltage and current time series, which are used to train  models described in this paper and make predictions, unfold in time and their temporal component is dominant. 
Therefore, we have decided to use \glspl{RNN} and employ their most efficient architectures, 
namely \gls{LSTM} and \gls{GRU} \cite{graves2012supervised, morton2016analysis, pouladi2015recurrent, chen2016efficient}.

\subsection{\glsentryshort{LSTM}}

\begin{figure}
	\begin{subfigure}[t]{0.48\hsize}
	    \centering
		\includegraphics[width=0.8\hsize]{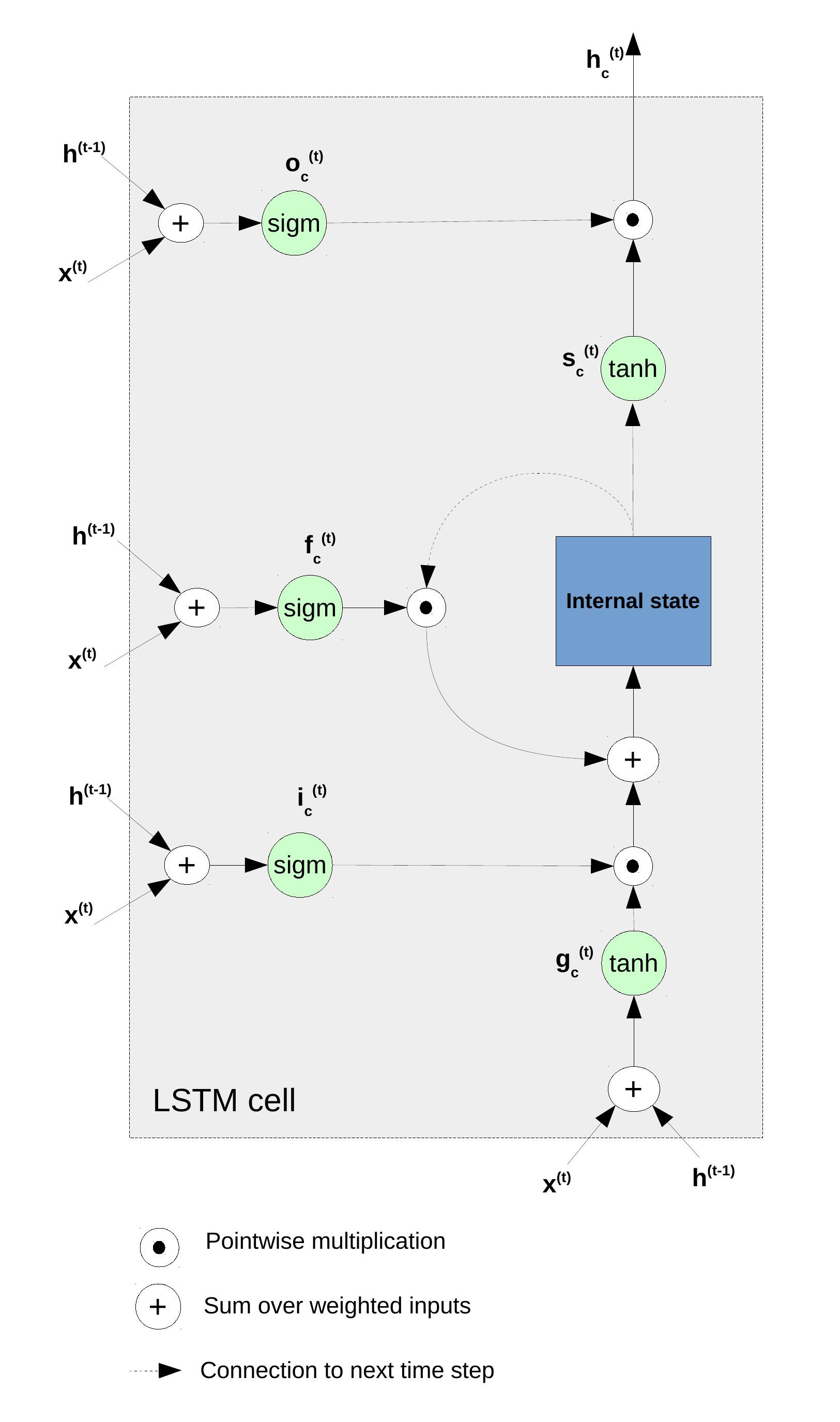}
		\caption{}
		\label{fig:lstm_block_architecture}
	\end{subfigure}
	\begin{subfigure}[t]{0.48\hsize}
	    \centering
		\includegraphics[width=0.8\hsize]{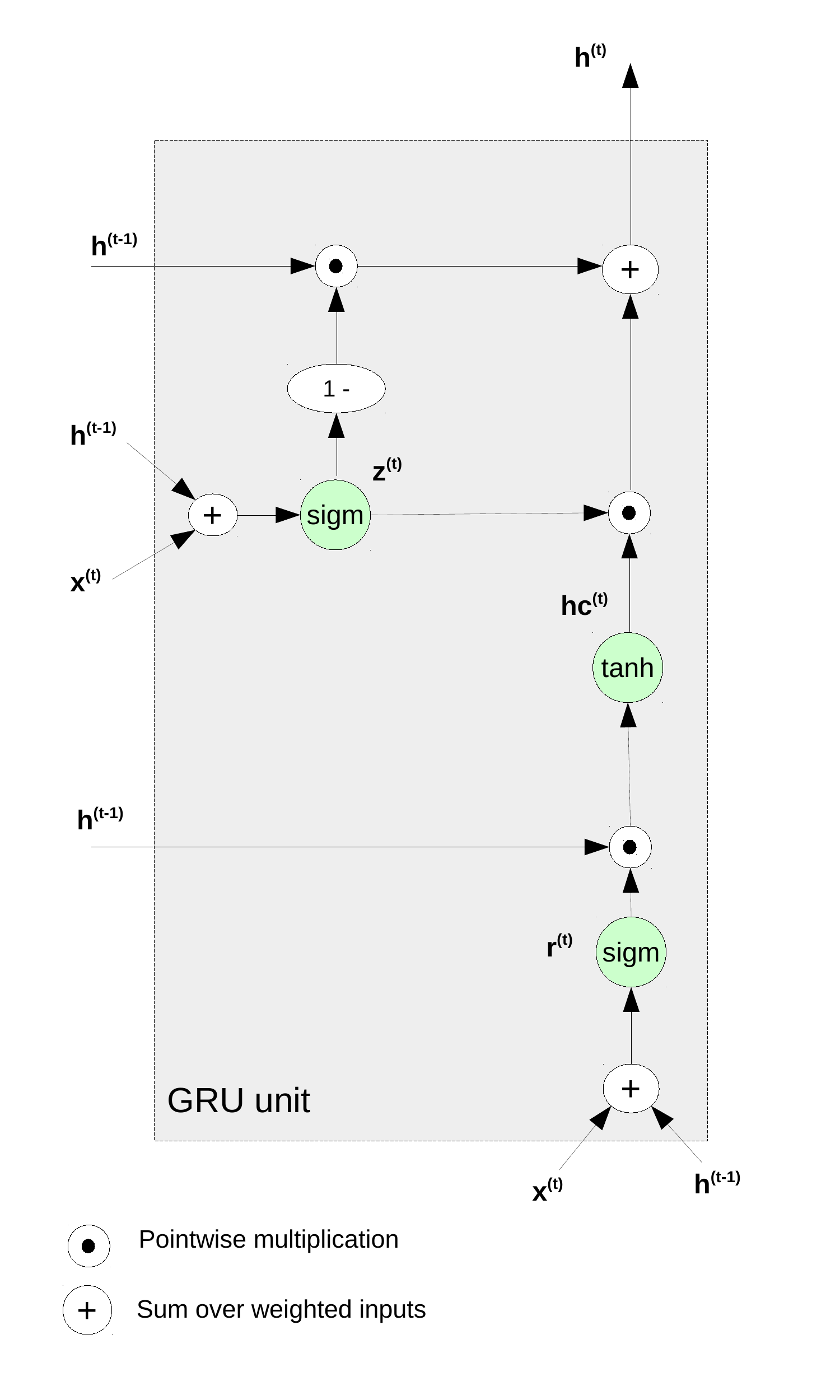}
		\caption{}
		\label{fig:gru_architecture}
	\end{subfigure}
	\caption{An architecture of the \subref{fig:lstm_block_architecture} LSTM cell and \subref{fig:gru_architecture} GRU unit}
\end{figure}

\glsreset{LSTM}

The \gls{LSTM} internal structure is based on a set of connected cells. The structure of a single cell is presented in Fig.~\ref{fig:lstm_block_architecture}. It contains feedback connection storing the temporal state of the cell, three gates and two nodes which serve as an interface for information propagation within the network \cite{zachary2015critical}. 

There are three different gates in each LSTM cell:
\begin{itemize}
 \item \textit{input} gate $i_c^{(t)}$ which controls input activations into the memory element,
 \item \textit{output} gate $o_c^{(t)}$ controls cell outflow of activations into the rest of the network,
 \item \textit{forget} gate $f_c^{(t)}$ scales the internal state of the cell before summing it with the input through the self-recurrent connection of the cell. This enables gradual forgetting in the cell memory.
\end{itemize}
\noindent
In addition, the LSTM cell also comprises of an input node $g_c^{(t)}$ and an internal state node $s_c^{(t)}$.

Modern \gls{LSTM} architectures may also contain \textit{peephole connections} \cite{greff2015lstm}. Since they are not used in the experiment, they were neither depicted in Fig.~\ref{fig:lstm_block_architecture} nor addressed in this description.

The output of a set of \gls{LSTM} cells is calculated according to the following set of vector equations:

\begin{equation}
g^{(t)} = \phi(W_{gx}x^{(t)} + W_{gh}h^{(t-1)} + b_g),
 \label{eq:lstm_input_node}
\end{equation}

\begin{equation}
i^{(t)} = \sigma(W_{ix}x^{(t)} + W_{ih}h^{(t-1)} + b_i),
 \label{eq:lstm_input_gate}
\end{equation}

\begin{equation}
f^{(t)} = \sigma(W_{fx}x^{(t)} + W_{fh}h^{(t-1)} + b_f),
 \label{eq:lstm_forget_gate}
\end{equation}

\begin{equation}
o^{(t)} = \sigma(W_{ox}x^{(t)} + W_{oh}h^{(t-1)} + b_o)
 \label{eq:lstm_output_gate}
\end{equation}

\begin{equation}
s^{(t)} = g^{(t)} \odot i^{(t)} + s^{(t-1)} \odot f^{(t)},
 \label{eq:lstm_internal_state}
\end{equation}

\begin{equation}
h^{(t)} = \phi (s^{(t)}) \odot o^{(t)}.
 \label{eq:lstm_output_node}
\end{equation}

While examining (\ref{eq:lstm_input_node}) -- (\ref{eq:lstm_output_node}), it may be noted that instances for a current and previous time step are used in the calculation of the output vector of hidden layer $h$ as well as for the internal state vector $s$. Consequently, $h^{(t)}$ denotes a value of an output vector at the current time step, where as $h^{(t-1)}$ refers to the previous step. It is also worth noting that the equations contain vector notation which means that they address the whole set of \gls{LSTM} cells. In order to address a single cell a subscript $c$ is used as it is presented in Fig.~\ref{fig:lstm_block_architecture}, where for instance $h_c^{(t)}$ refers to a scalar value of an output of this particular cell.

The \gls{LSTM} network learns when to let an activation into the internal states of its cells and when to let an activation of the outputs. This is a gating mechanism and all the gates are considered as separate components of the \gls{LSTM} cell with their own learning capability. This means that the cells adapt during training process to preserve a proper information flow throughout the network as separate units. This means that when the gates are closed, the internal cell state is not affected. In order to make this possible a hard sigmoid function $\sigma$ was used, which can output \num{0} and \num{1} as given by (\ref{eq:hard_sigma}). This means that the gates can be fully opened or fully closed.

\begin{equation}
\begin{split}
\sigma (x) =
\begin{cases}  
0 \text{~if~}  x \leq t_{l}, \\ 
ax + b  \text{~if~} x \in (t_{l}, t_{h}), \\
1 \text{~if~}  x \geq t_{h}. \\ 
\end{cases}
\label{eq:hard_sigma}
\end{split}
\end{equation}

In terms of the backward pass, a so-called constant error carousel enables the gradient to propagate back through many time steps, neither exploding nor vanishing \cite{hochreiter1997long, zachary2015critical}. 

\subsection{\glsentryshort{GRU}}

\glsreset{GRU}

Since its invention in 1997, the \gls{LSTM} was updated and modified \cite{greff2015lstm} to improve its modeling properties and reduce large computational demands of the algorithm. It is worth noting that \gls{LSTM}, as opposed to a vanilla \gls{RNN}  \cite{wielgosz2016usingLSTM} is much more complex in terms of the internal component constituting its cell. This results in a long training time of the algorithm. Therefore there were many experiments conducted with simpler architectures which preserve beneficial properties of \gls{LSTM}. One of such algorithms is the \gls{GRU} \cite{chung2015gated} which is widely used in \glsentrylong{DL} as an alternative for \gls{LSTM}. According to the recent research results it even surpasses \gls{LSTM} in many applications \cite{chung2014empirical}. 

\gls{GRU} has gating components which modulate the flow of information within the unit as presented in Fig.~\ref{fig:gru_architecture}.

\begin{equation}
h^{(t)} = (1 - z^{(t)}) \odot h^{(t-1)} + z^{(t)} \odot hc^{(t)},
 \label{eq:gru_h}
\end{equation}

The activation of the model at a given time $t$ is a linear interpolation between the activation $h^{(t-1)}$ from the previous time step and the candidate activation $hc^{(t)}$. The activation is strongly modulated by $z^{(t)}$ as given by (\ref{eq:gru_h_unfolded_0}) and  (\ref{eq:gru_h_unfolded_1}).

\begin{equation}
h^{(t)} = (1 - z^{(t)}) \odot h^{(t-1)} + z^{(t)} \odot hc^{(t)} = h^{(t-1)} - z^{(t)} \odot h^{(t-1)} + z^{(t)} \odot hc^{(t)},
 \label{eq:gru_h_unfolded_0}
\end{equation}

\begin{equation}
h^{(t)} =  h^{(t-1)} - z^{(t)} \odot (h^{(t-1)} + hc^{(t)}),
 \label{eq:gru_h_unfolded_1}
\end{equation}

\begin{equation}
z^{(t)} = \sigma(W_{z}x^{(t)} + W_{z}h^{(t-1)}),
 \label{eq:gru_z}
\end{equation}

Formula for the update gate is given by (\ref{eq:gru_z}) and modulates a degree to which a GRU unit updates its activation. 
The GRU has no mechanism to control to what extent its state is exposed, but it exposes the whole state each time. 

\begin{equation}
r^{(t)} = \sigma(W_{r}x^{(t)} + W_{r}h^{(t-1)}),
 \label{eq:gru_r}
\end{equation}

The response of the reset gate is computed according to the same principle as the update gate. Previous state information is multiplied by the coefficients matrix $W_{r}$ and so is the input data. It is computed by (\ref{eq:gru_r}).

\begin{equation}
hc^{(t)} = \phi(Wr^{(t)} \odot h^{(t-1)} + Wx^{(t)}),
 \label{eq:gru_hc}
\end{equation}

The candidate activation $hc^{(t)}$ is computed according to  (\ref{eq:gru_hc}). When $r^{(t)}$ is close to 0, meaning that the gate is almost off, the stored state is forgotten. The input data is read instead.

\section{Operation layer}
\label{section:operation_layer}

This section briefly discuss an architecture of a system protecting \gls{LHC} against equipment failures with special emphasis to software system dedicated to collection and analysis 
of data recorded at a time of failure. A set of data extracted from the data acquired within \gls{LHC} protection system was used as 
a learning dataset for experiments described in \ref{subsection:dataset}. 

\subsection{The \glsentryshort{LHC} \glsentrylong{MPS}}

The \gls{LHC} is an experimental device composed of hundreds of modules which constitute a large system. 
The tunnel and the accelerator is just a very critical tiny fraction of the \gls{LHC} infrastructure.  
The energy stored in the superconducting circuit of main magnets of each sector of the \gls{LHC} at \SI{13}{\kilo\ampere} amounts to about \SI{1.2}{\giga\joule}, sufficient to
heat up and melt \SI{1900}{\kilogram} of copper. At \SI{7}{\tera\electronvolt} each proton beam accumulates an energy of \SI{360}{\mega\joule}, equivalent to the
energy for warming up and melting \SI{515}{\kilogram} of copper. It is a hundred times higher than previously achieved in any accelerator. 
Therefore the machine must be protected against consequences of malfunction of almost each its element. 
An energy corresponding to a fraction of some \num[retain-unity-mantissa = false]{1e-7} of the beam energy can quench a dipole magnet when operates at full current. 
The critical safety levels are therefore required to operate the \gls{LHC}.
A system dedicated to fulfill this important role is known as \gls{MPS} \cite{MPS_Wenninger, interlocks, MPS_Schmidt}. In general it consists of two interlock systems: the \gls{PIS} and the \gls{BIS}. 
The \gls{BIS} is a superordinate system which collects signals from many sources. There are currently \num{189} inputs from client systems. We can distiguished several sources:
\begin{itemize}
 \item the \gls{BLM};
 \item the \gls{BPM};
 \item the \gls{WIC};
 \item the \gls{FMCM};
 \item the collimation system;
 \item the personnel access system;
 \item the operator inhibit buttons;
 \item the vacuum valves;
 \item the interlock signals from the experiments.
\end{itemize}
However the most important and the most complex protection subsystem is the \gls{PIS} which ensures communication between systems involved in the powering of the \gls{LHC} superconducting magnets. 
This includes the \gls{PC}, the \gls{QPS}, the \gls{UPS}, the \gls{AUG} and the cryogenic system. When a magnet quench is detected by the \gls{QPS}, the power
converter is turned off immediately. In total, there are order of  thousands of interlock signals. The signals are distributed mainly by three different arrangements:
\begin{itemize}
 \item point to point connections with one source and one receiver;
 \item field-bus is used to create a software-based link in less critical cases, in particular to give permission for powering etc.;
 \item current loops which are used to connect many sources to several receivers.
\end{itemize}
A current loop is a current source with a large compliance which force a constant current through a line connecting reed relays or solid-state switches (opto-couplers) 
installed in each module along whole \gls{LHC} sector.
A request for termination of the operation of the whole machine is triggered by opening one switch in the line. 
The interruption of the current generates a trigger signal of the interlock controller.

When a failure is detected that risks stopping the powering of magnets, a beam dump request is sent to the \gls{BIS}. It generates three signals. A first is sent to
the \gls{LBDS} to request the extraction of the beams. A second signal is sent to the injection system to block injection into the \gls{LHC} as well as extraction of beam from the \gls{SPS}.
A third signal is a trigger for the timing system that sends out a request to many \gls{LHC} systems for providing data that were recorded before the beam dump, to understand the reasons for 
the beam dump. A device in these kind of systems comprises a circular buffer which at any time contains current information about the protected component. In particular case of a quench detector,  
the buffer contains voltage time series acquired with a high resolution time by an ADC connected to a superconducting coil. At a trigger time the half of the buffer space is already filled with samples 
acquired before an event (quench) time. After an event time the voltage samples are still recorded to fill the rest of the buffer space. Therefore the buffer contains time series around 
trigger time at both sides. This kind of data is called ``post-mortem'' because it is recorded after the component ceased its regular activity due to a malfunction.

\begin{figure}
\centering
{%
\setlength{\fboxsep}{3pt}%
\setlength{\fboxrule}{1pt}%
\fbox{\includegraphics[width=0.8\hsize]{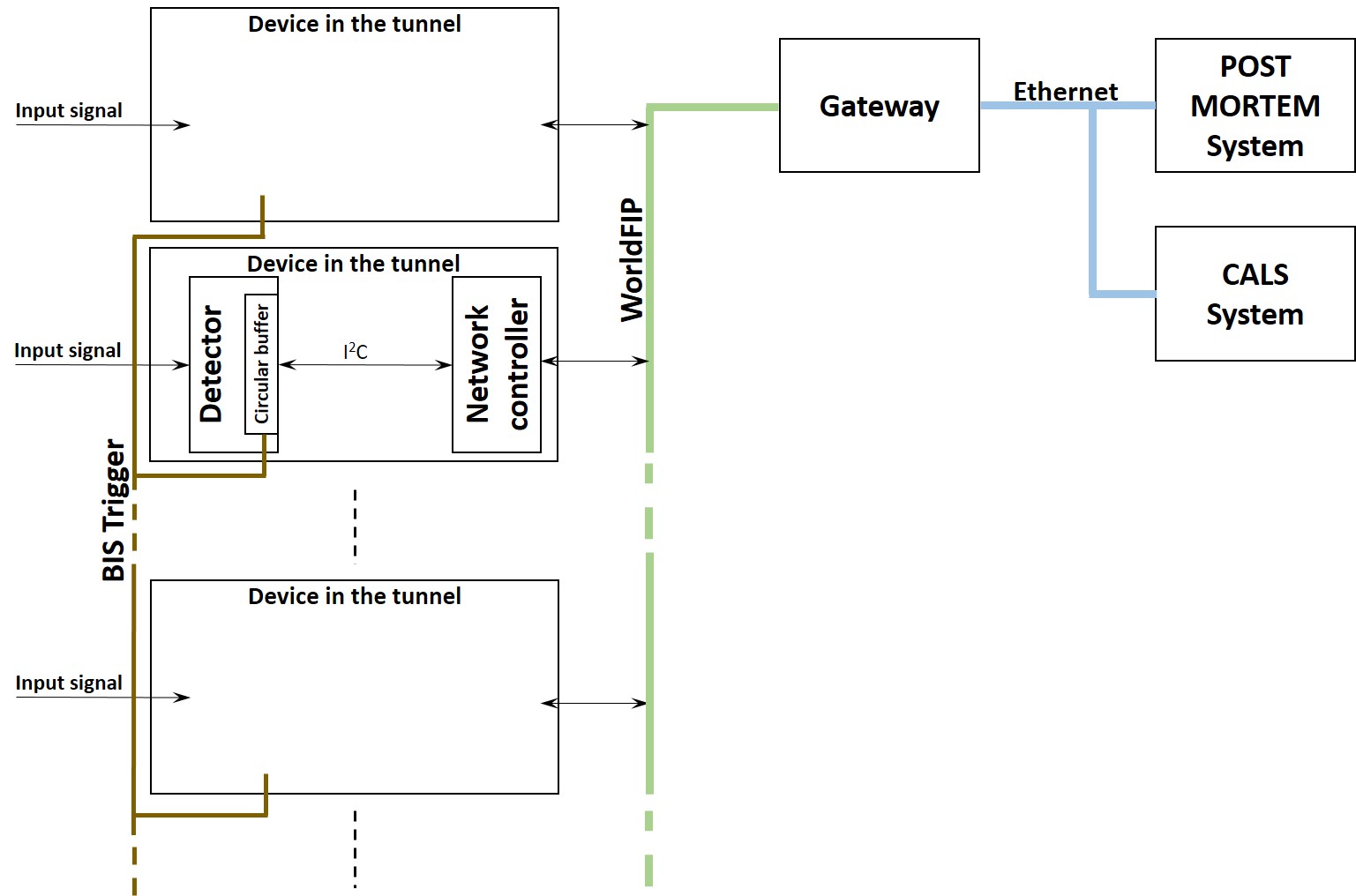}}
}%
\caption{The hardware path of signals from individual devices in the tunnel to the \glsentrylong{PMS}. 
The green line marks a kind of industrial computer network protocol used for real-time distributed control called \gls{WorldFIP}. }
\label{fig:sig_hw_path}
\end{figure}

The contents of the buffer is sent out by the network controller of the device over the field-bus to a gateway. 
Next the data is transfered to a database over Ethernet network. The transmition's path of the data is shown on Fig.~\ref{fig:sig_hw_path}. 
There are two arrival points for data. Both are huge software systems to store and process data about any \gls{LHC} module. 
First system is used during failures and requested checks. It is described below in \ref{subsection:PMS}.
The Fig.~\ref{fig:sig_hw_path} includes also a second system for permanent acquiring of equipment data. The \gls{CALS} is used to store and retrieve billions of data records per day, 
from across the complete CERN accelerator complex, related subsystems, and experiments. It is not a subject of this description. 

\subsection{The \glsentryshort{LHC} \glsentrylong{PMS}}
\label{subsection:PMS}

The \gls{PMS} is a diagnostics tool with the role of organizing the collection and analysis of transient data recorded during time interval around a failure or a request sent by any device 
in the \gls{MPS} \cite{Ciapala:691828}. The main purpose is to provide a fast and reliable tool for the equipment experts and the operation crews 
to help them decide whether accelerator operation can continue safely or whether an intervention is required.
The most important parameters from \gls{LHC} systems are stored in circular buffers inside the individual devices. The aim is to process the contents of the buffers after an event i.e. \textit{post mortem}.
When a failure (a beam loss or a magnet quench) happens, a trigger is generated by the \gls{BIS}. The buffers are then frozen and transmitted to the \gls{PMS} for further storage 
and analysis \cite{Ciapala:691828, Lauckner:567214, Borland:1998}. The transmission is undertaken by the controllers of the equipment that send the data at the arrival of a trigger. 
The hardware path of signals stored in \gls{PMS} is presented in the Fig.~\ref{fig:sig_hw_path}. 

\begin{figure}
\centering
{%
\setlength{\fboxsep}{3pt}%
\setlength{\fboxrule}{1pt}%
\fbox{\includegraphics[width=0.8\hsize]{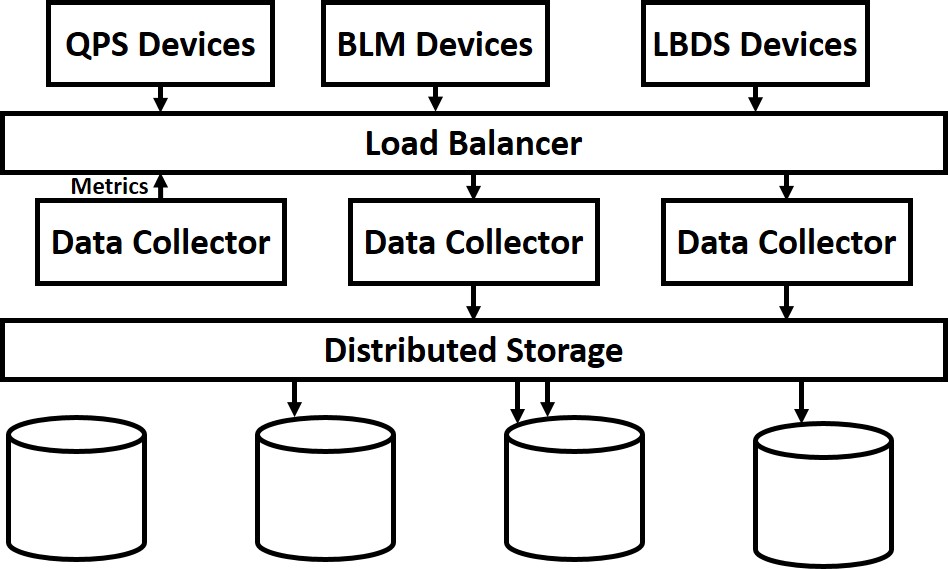}}%
}%
\caption{The simplified \glsentrylong{PM} architecture \cite{PMmigration}.}
\label{fig:pm_arch}
\end{figure}

When implementing such a system, a number of challenges to overcome arises. The devices are distributed over the entire ring and therefore a correct synchronization and 
a precise time-stamping at the equipment level is necessary to reconstruct the event development. The value of parameters like buffer depth and sampling rate must be considered for 
each kind of devices separately. The solution of the \gls{PMS} was modified and developed during hardware commissioning, first experimental run and \gls{LS1}. The current architecture is presented 
in the Fig.~\ref{fig:pm_arch}. It provides a scalability both in vertical and in horizontal directions. The vertical scalability means that resources can be added to the nodes with a minimum downtime and impact on the 
service availability. The horizontal scalability is provided using three features. The first feature is a dynamic load distribution during data collecting. Any device can dump the \gls{PM} data to any Data 
Collector transparently and without any additional configuration effort. This way the load can be distributed among the Data Collectors. The second feature is a data storage redundancy. 
The Data Collector that processes the dump writes the data to the distributed storage. The data are automatically replicated by the storage infrastructure. Third feature is a data consistency 
check. The storage infrastructure provides also an integrity verification and a detection and correction of errors. 

A method of serialization of PM data has to ensure:
\begin{itemize}
  \item data splittabilty because a user usually runs an analysis only on a part of data dump,
  \item data compression because the signals often contain zeros and an optimization of the occupied space in the storage system is desired.
\end{itemize}
The technology called Apache Avro\texttrademark\ was choosen. Avro brings the flexibility required for the data structure. Avro relies on schemas. When Avro's data is stored in a file, 
its schema is stored with it, so that files may be processed later by any program. When Avro's data is read, the schema which was applied when writing is always present. Therefore Avro is a 
self-describing format. Avro's schemas are defined with \gls{JSON}. This facilitates implementation in languages that already have \gls{JSON} libraries.

The data transfer from devices relies entirely on the \gls{CMW} \gls{RDA} protocol \cite{CMW, CMWnew}. 
The main goal of \gls{CMW} is to unify a middle layer used to build every control system for operarion of accelerators at \gls{CERN}.
Currently data collection uses RDA2 based on CORBA (old) and RDA3 based on ZeroMQ (new). 

Users can access to the \gls{PM} data by means of using a specially designed \gls{API}. This \gls{API} was designed using software architecture called \gls{REST}. 
The aim is to serve multiple language technologies according to user preferences: Python, MATLAB, LabVIEW, C++ and Java. A user is not dependent on the data format and the file system. 
A direct extraction of only one signal from a big dataset is possible without necessity of reading the entire set. The \gls{API} can handle very complex queries.

\subsection{The \glsentryshort{LHC} \glsentrylong{PMAF}}

\begin{figure}
\centering
\includegraphics[width=0.8\hsize]{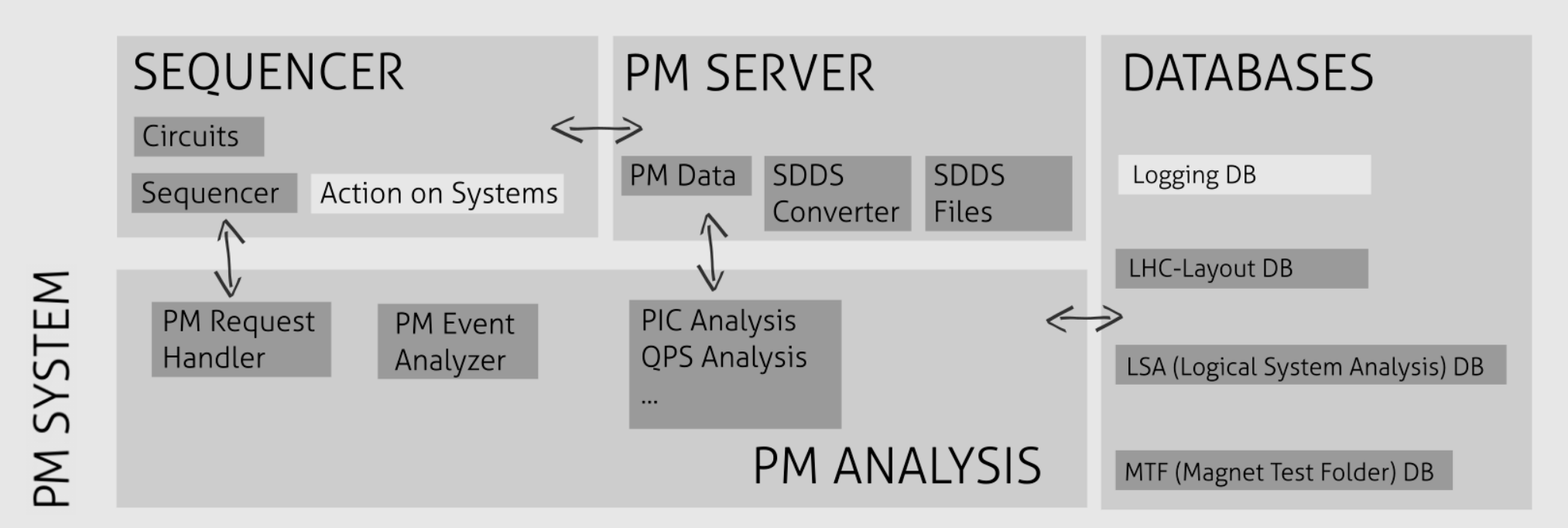}
\caption{The building blocks of the \glsentrylong{PMS}}
\label{fig:post_mortem_system}
\end{figure}

In the Fig.~\ref{fig:post_mortem_system} building blocks of the \gls{PMS} that is surrounded by the sequencer and the databases with two main parts, the server and the client can be seen.

In case of a Hardware Commissioning the sequencer application controls the power converters. They execute a current cycle or a ramp. The \glsentrylong{PM} Request Handler combines the \gls{PM} data 
with the test performed by the sequencer and the \glsentrylong{PM} Event Analyser collects all such events for the presentation and subsequent analysis to 
the equipment experts. The Analyser allows the experts to execute different analysis programs and data viewers. With them they can verify the success of the test and use an electronic 
signature to pass or fail it. The final result is being sent to the sequencer for upload into the \gls{MTF} database. There, a decision is made either to accept the test, to repeat the test or 
to open a procedure for non-conformity. Different analysis programs and data viewers (service) have been developed on the \gls{PMS}.

The \gls{PM} service has been providing data collection, storage and analysis of \gls{LHC} event data since 2008. Around \num{20} different client systems are today sending data to the \gls{PM} 
servers, in case of beam dumps as much as \num{3000} individual files (containing up to \SI{50}{\giga\byte} of data) in a period of less than a few \num{10} seconds \cite{Andreassen:1235888}.

\begin{figure}
\centering
\includegraphics[width=0.8\hsize]{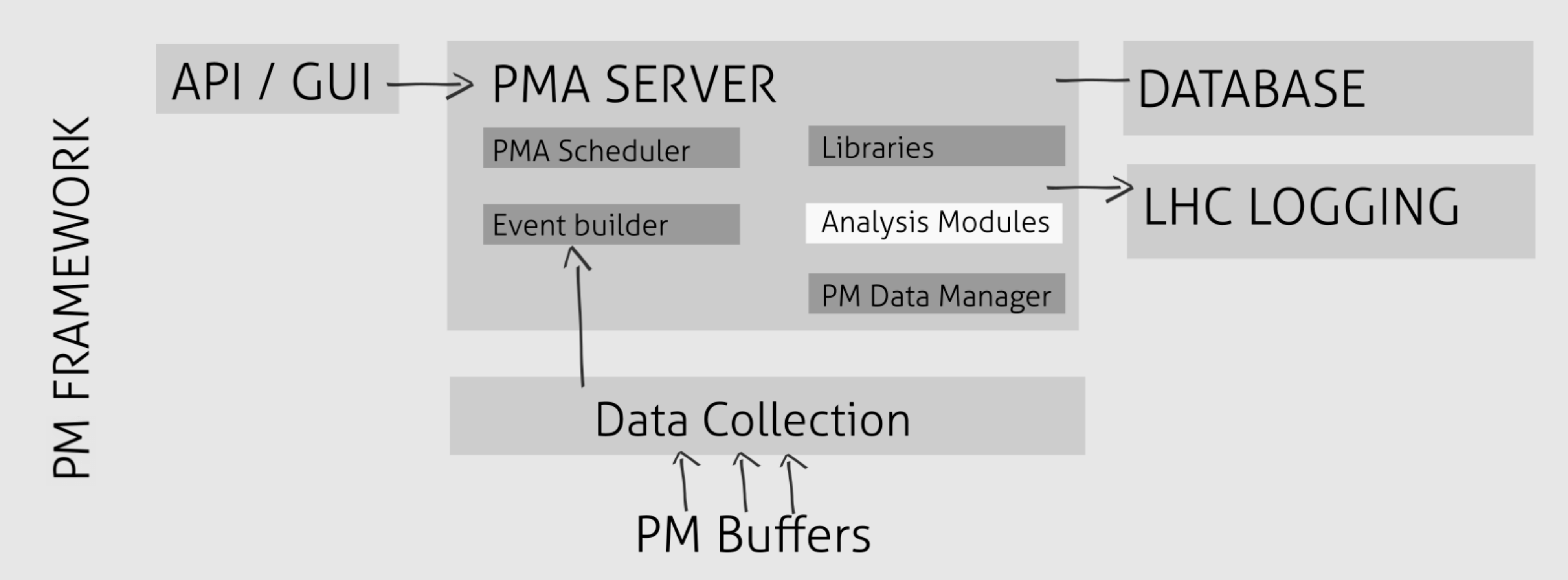}
\caption{The simplified digaram of the main components of the \glsentrylong{PMAF}.}
\label{fig:post_mortem_framework}
\end{figure}

Analysis of these transient data requires an efficient and accurate analysis for the thousands of \gls{PM} data buffers arriving at the system in the case of beam dumps. 
The \gls{LHC} \glsentrylong{PMF} orchestrates the analysis flow and provides all necessary infrastructure to the analysis modules such as read/write APIs for \gls{PM} data, 
database/reference access, analysis configurations, etc. Fig.~\ref{fig:post_mortem_framework} presents main parts of \gls{PMF}.

The key component of the \gls{PMAF} is an Event Builder. This application detects interesting sets of \gls{PM} data which subsequently become the subject of a detailed analysis by different 
Analysis Modules. Modules are prepared taking into account a domain knowledge related to specific class of equipment. 

\subsection{Anomaly detection and \glsentrylong{PM}}

\begin{figure}
\centering
\includegraphics[width=0.8\hsize]{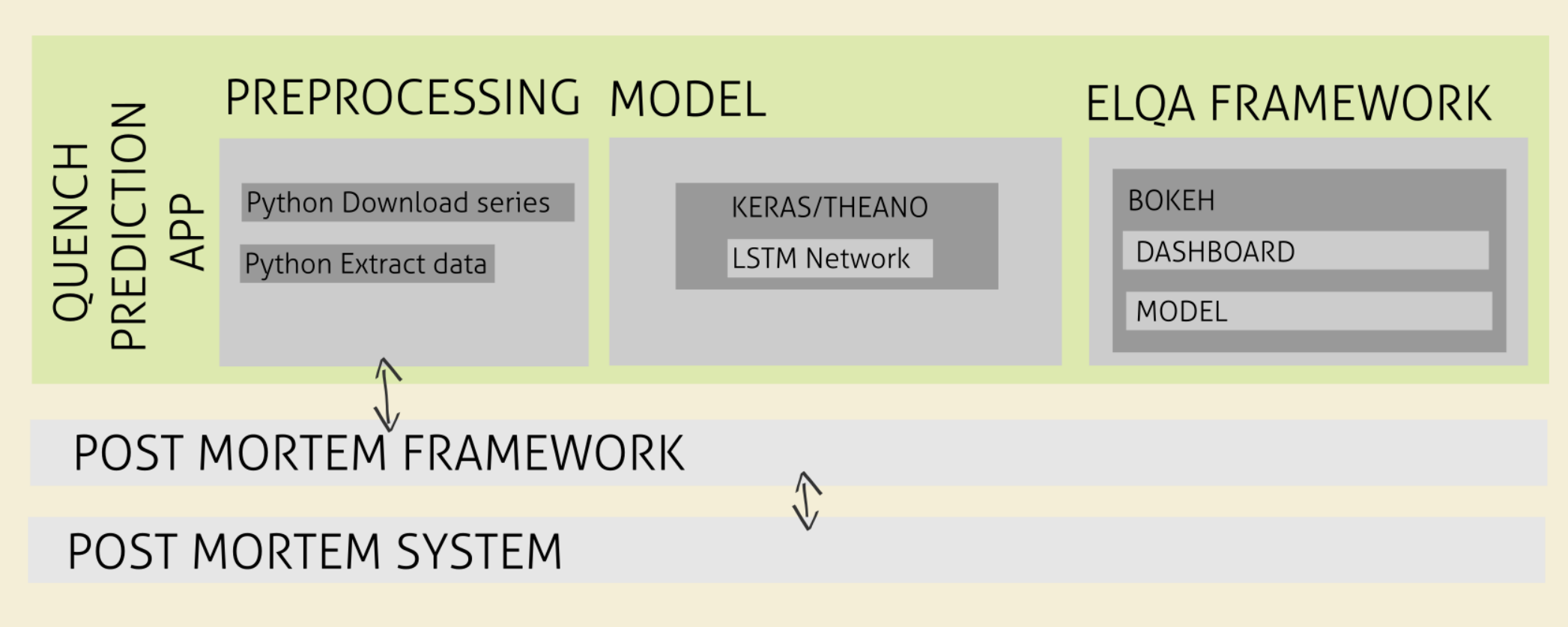}
\caption{The main components of the software for \glsentrylong{DL} research employed in this work.}
\label{fig:main_components_deep_learning}
\end{figure}

In our research, we used \gls{PM} \gls{JSON} \gls{API} written in Python to gather targeted data for an anomaly detection/prediction. 
Customized preprocessors were developed to access the framework in order to generate learning dataset. A \glsentrylong{DL} model was build with the Keras/Theano libraries \cite{chollet2015}, 
where we use an \gls{LSTM}/\gls{GRU} model as described in the subsection \ref{subsection:dataset}. 
Fig.~\ref{fig:main_components_deep_learning} presents a schema for the experiment with its key components.

The presentation of the results of the model are intended to be integrated in a web application for quench detection. For this purpose \gls{ELQA} framework, developed at 
\glsentryshort{TE-MPE-EE}, will be used \cite{articleELQA,mertikdhalerup}. \gls{ELQA} framework is a framework for developement of interactive web applications for data analysis. 
It supports integration of various generated machine learning models with graphical user interfaces within a browser in an efficient way. It is developed in Python with opensource libraries 
such as Scikit-learn for machine learning and Bokeh, an interactive visualization library that targets modern web browsers for presentation \cite{Bokeh2016}.

\section{Proposed method for anomaly detection}
\label{section:proposed_method}

\begin{figure}
	\begin{subfigure}{\hsize}
		\centering
		\includegraphics[width=0.9\hsize]{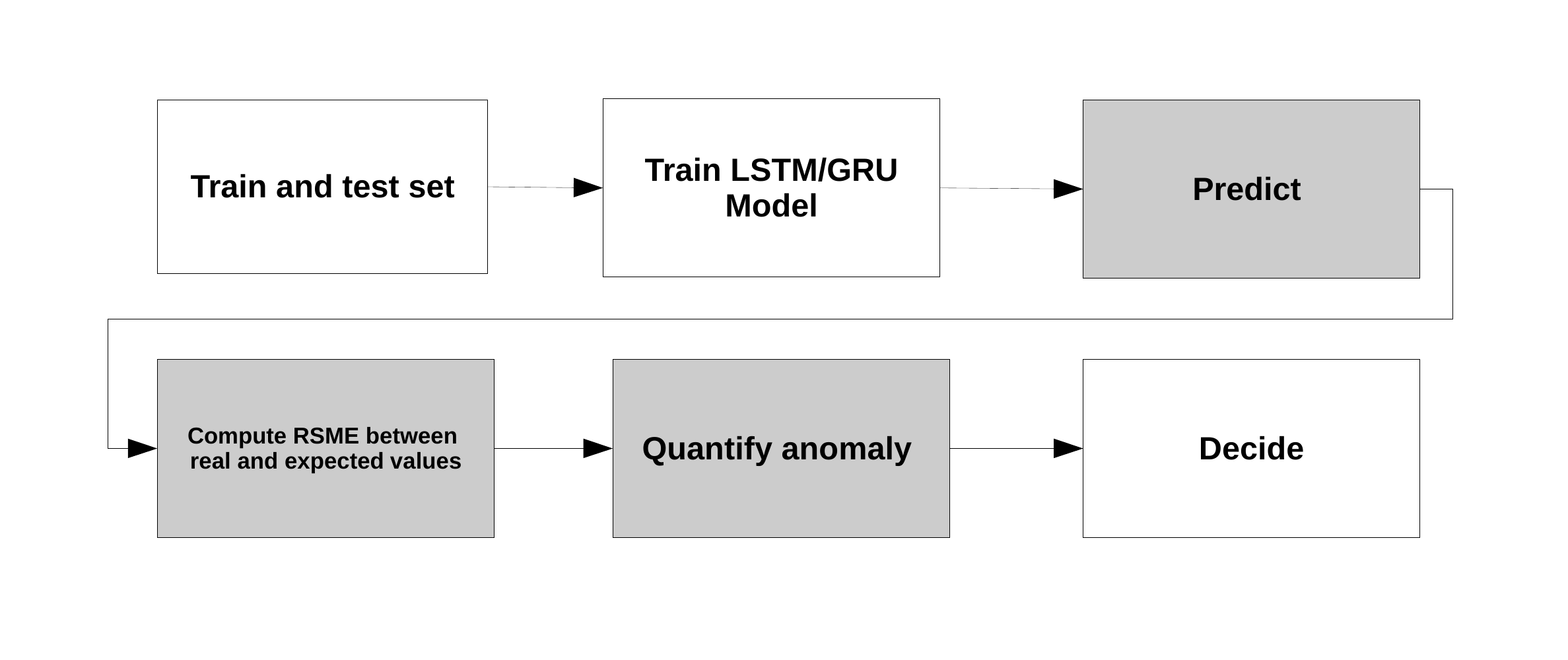}
		\vspace{-2.5em}
		\caption{}
		\label{fig:setup_rmse_prediction}
	\end{subfigure}
	\begin{subfigure}{\hsize}
		\centering
		\includegraphics[width=0.9\hsize]{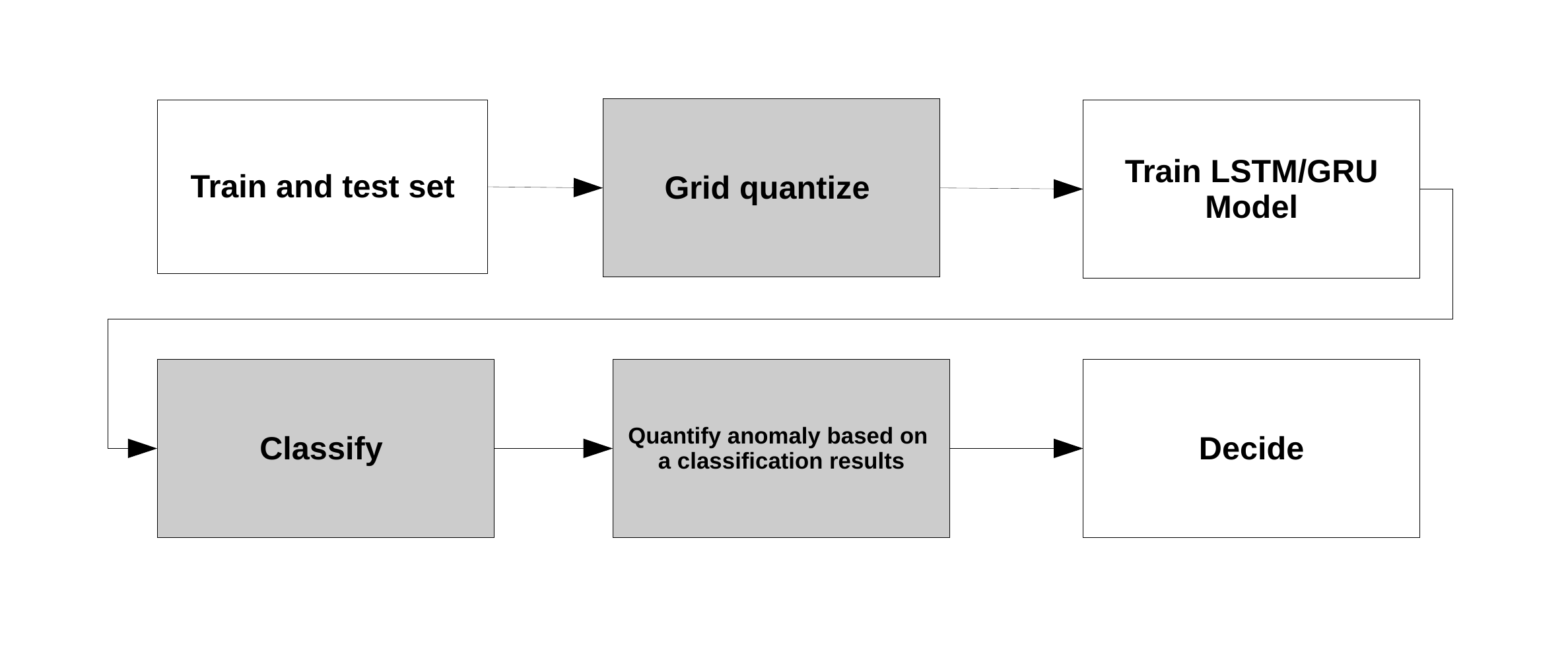}
		\vspace{-2.5em}
		\caption{}
		\label{fig:setup_grid_classification}
	\end{subfigure}
	\caption{Experimental setups featuring \subref{fig:setup_rmse_prediction} \gls{RMSE} and prediction and \subref{fig:setup_grid_classification} grid quantization and classification}
\end{figure}

In \cite{wielgosz2016usingLSTM}, the experiments with the Timber database (Timber is the user interface to the LHC Logging System) were conducted using the setup presented in Fig.~\ref{fig:setup_rmse_prediction}, which employed \gls{RMSE} measure for anomaly detection. A huge challenge in this approach is a lack of a clear reference threshold of an anomaly. In order to determine the error level, a group of experts must be consulted and it is not always easy to set one. This is due the fact that \gls{RMSE} does not always indicate anomalous behavior well enough to quantify it correctly \cite{strecht2015comperative}.

We decided to take advantage of the experience from \cite{wielgosz2016usingLSTM} and introduced a new experimental setup which is shown in Fig. \ref{fig:setup_grid_classification}. This new approach allowed to convert a regression task to the classification one, which in turn enables better anomaly quantification.

The main difference between the previously used approach and the proposed one is an introduction of a grid quantization and classification steps (see marked boxes in both Fig.~\ref{fig:setup_rmse_prediction} and \ref{fig:setup_grid_classification}). Consequently, in the new approach the train and test data are brought to several categories depending on a grid size. This transformation may be perceived as a specific kind of quantization, since the floating-point data are converted to the fixed-point representation denoted as categories in this particular setup. It is worth noting that increase in the grid size leads to an increase of the resolution and it is more challenging for the classifier. Potentially, large resolution setup will demand larger model.

\begin{figure}
\centering
\includegraphics[width=0.49\hsize]{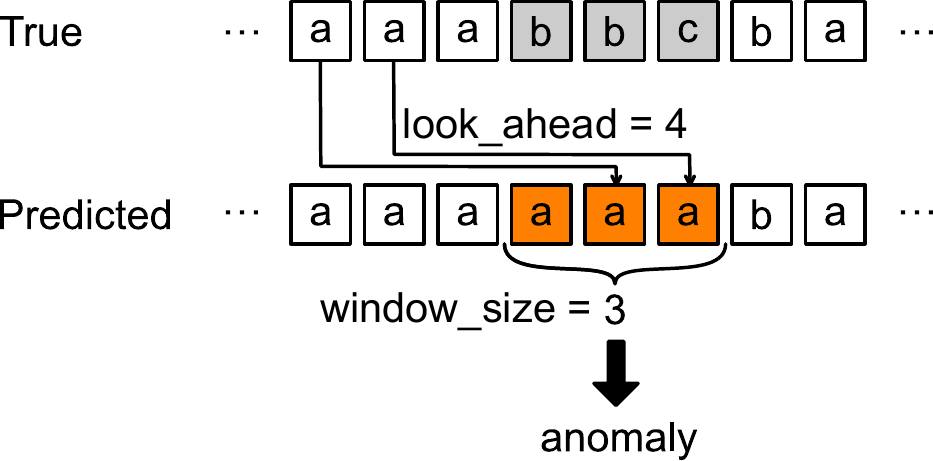}
\caption{Visualization of the proposed method}
\label{fig:method_visualization}
\end{figure}

Introduction of the grid quantization guaranties maximum error rate within each category. For instance, if the grid size is \num{10}, the guaranteed maximum error is \SI{10}{\percent} according to the accuracy quality measure. Once the grid size is increased to \num{20}, the guaranteed maximum error is \SI{5}{\percent}. In order to determine if an anomaly occurs it is enough to observe the error level for several time steps. When it turns out that over this time period the error exceeds \SI{5}{\percent} for the grid size of \num{20}, it means that the anomaly occurred (Fig.~\ref{fig:method_visualization}). The data expert has a much easier task in this case, because the only decisions required are about the grid size and the anomaly detection window, both of which are well quantifiable parameters.

It is worth emphasizing that the proposed approach is based on an assumption that a very well trained model is used. Its performance should be in a range between \num{90} and \SI{100}{\percent} in terms of accuracy. This is a foundation of choosing a reliable anomaly detection window. 

The anomaly detection window is a parameter that determines how many consecutive predicted values in the signal need to differ from the true ones in order to detect an anomaly. Each predicted value that matches a true one resets the difference counter. A small anomaly detection window allows for a faster reaction time, while bigger one decreases the possibility of a false positive.

The anomaly detection window size is related to the look\_ahead parameter of the model (how many time steps into the future model predicts) ie. look\_ahead value must be bigger than the window size. Such a condition is necessary in order to avoid the influence a possible anomaly could have on values predicted within the window.
\section{Experiments and the discussion} 
\label{section:experiments}

A main goal of the conducted experiments was validation of the feasibility of the application of the proposed method for detecting anomalies in \gls{PM} time series of \gls{LHC} superconducting magnets. It is worth noting that this approach may also be adopted to other applications of the similar profile.

\subsection{Dataset}
\label{subsection:dataset}

\begin{figure}
\centering
\includegraphics[width=0.9\hsize]{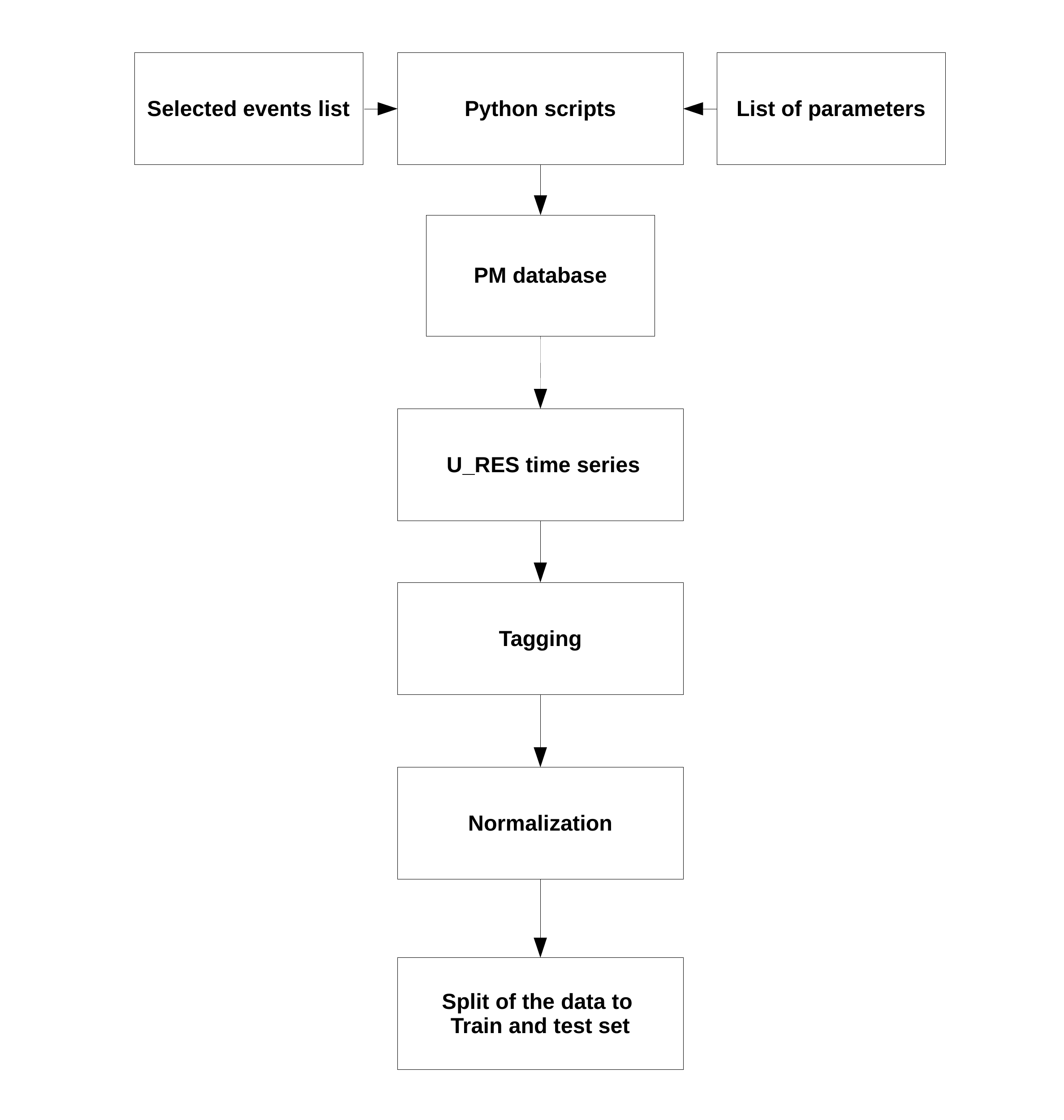}
\caption{The procedure for extraction of voltage time series with selected events from the \gls{PM} database using $U_{\mathit{res}}$ as an example}
\label{fig:getting_data_from_pm_database}
\end{figure}

All the data used for the experiments were collected from \gls{CERN} \gls{PM} database. The database contains various kinds of data acquired during both regular and special operating periods of \gls{LHC}. Whenever something extraordinary, like a quench, happens, the data is being acquired and collected in the database. Additionally, twice a day data is acquired during ramp-up and ramp-down phase. We have collected signals from \SI{600}{\ampere} magnets current for different time series: $U_{\mathit{res}}$, $U_{\mathit{diff}}$, $I_{\mathit{did}}$ and $I_{\mathit{dcct}}$.

\begin{table}
\caption{Created datasets}\label{tab:datasets}
\centering
\begin{tabular}{cc}
\toprule
Name&Size\\
\midrule
Small&\SI{47}{\mega\byte}\\
Medium&\SI{184}{\mega\byte}\\
Large&\SI{302}{\mega\byte}\\
\bottomrule
\end{tabular}
\end{table}

A procedure of data extraction from the \gls{PM} database is composed of several steps as presented in Fig.~\ref{fig:getting_data_from_pm_database}. A dedicated application and a set of parameters such as signal name and a number of time steps was used. \Gls{PM} database API does not allow to acquire more than one-day long signal at once. Therefore, the scripts were designed to concatenate several separate days to form a single data stream used for the experiments.

In total \SI{4}{\giga\byte} of data was collected from the database. Only a fraction of the data contained valuable information for our experiment. Consequently, we have provided a script to extract this information end keep it in separate files. Then we have divided them into three different data sets: the small, the medium and the large one (Tab. \ref{tab:datasets}). Such a division allowed to adjust hyper-parameters of the model with the small dataset before using the two remaining ones. As final steps, the data from each dataset was normalized to $[0, 1]$ range and split into train and test sets.

It is worth noting that most of the experiments presented in the experiment section of the paper were done with the smallest dataset because computation time was more feasible. A few experiments were conducted with the largest dataset to examine improvement of the model performance as a consequence of using more data.

\subsection{Quality assessment measure}
\label{subsection:quality_assessment_measure}

Accuracy is used as a quality evaluation of the experiments results presented in this paper. Is is calculated as follows:

\begin{equation}
    \mathit{Accuracy} = \sum\limits_{i}^{n} \frac{Y_{\mathit{true}}^i = Y_{\mathit{predicted}}^i}{n},
    \label{eq:class_accuracy}
\end{equation}

where $Y_{\mathit{true}}^i$ and $Y_{\mathit{predicted}}^i$ are the true categories and ones predicted by the model, respectively. The mean accuracy rate is calculated across all the predictions within a dataset of $n$ cardinality.

\subsection{Results}
\label{subsection:results}

\begin{figure}
	\begin{subfigure}{\hsize}
	    \centering
		\includegraphics[width=0.8\hsize]{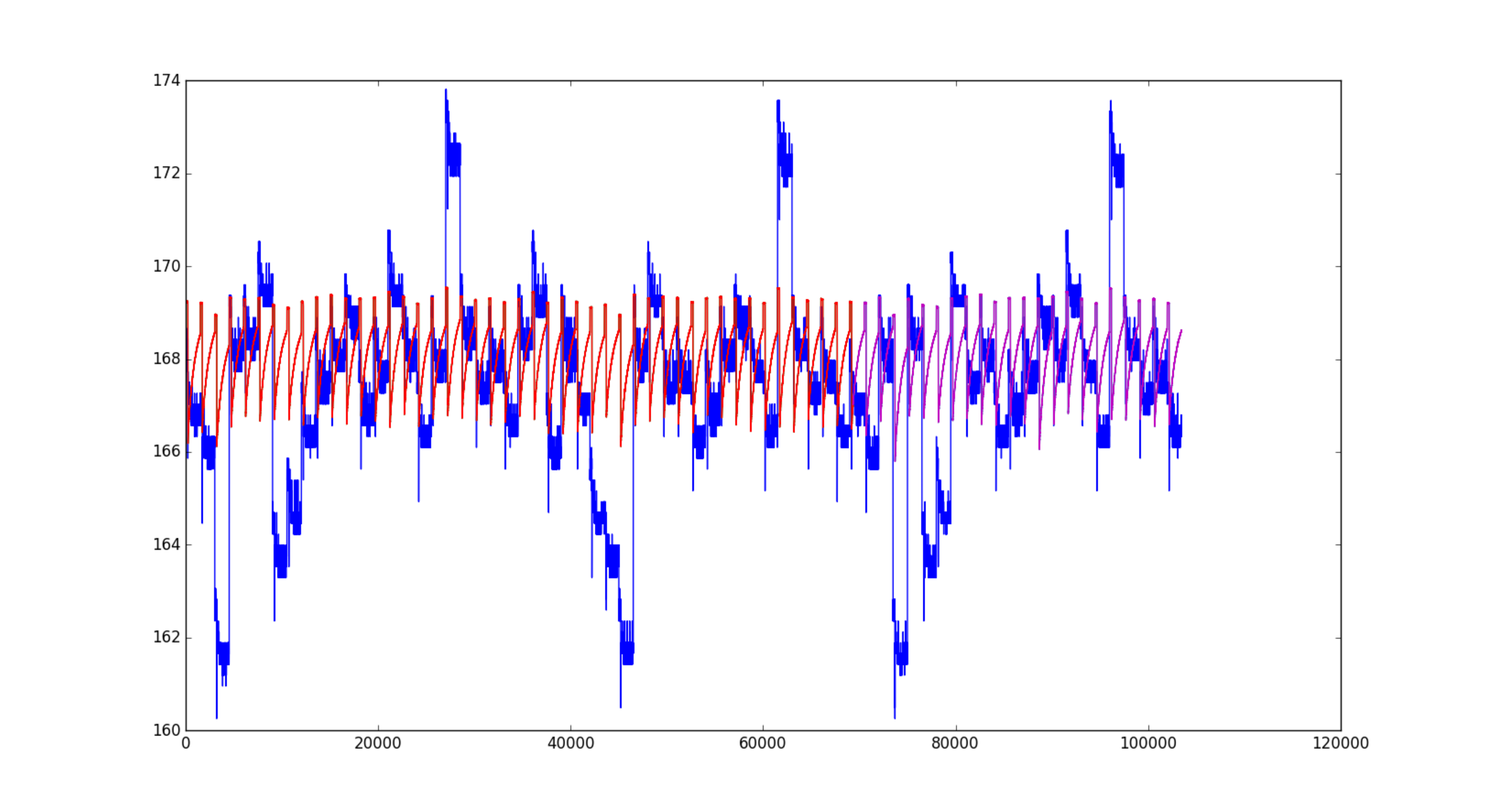}
		\caption{cells=1, epochs=2, RMSE = \SI{15.9}{\percent}}
		\label{fig:1_cell_2_epochs}
	\end{subfigure}
	\begin{subfigure}{\hsize}
	    \centering
		\includegraphics[width=0.8\hsize]{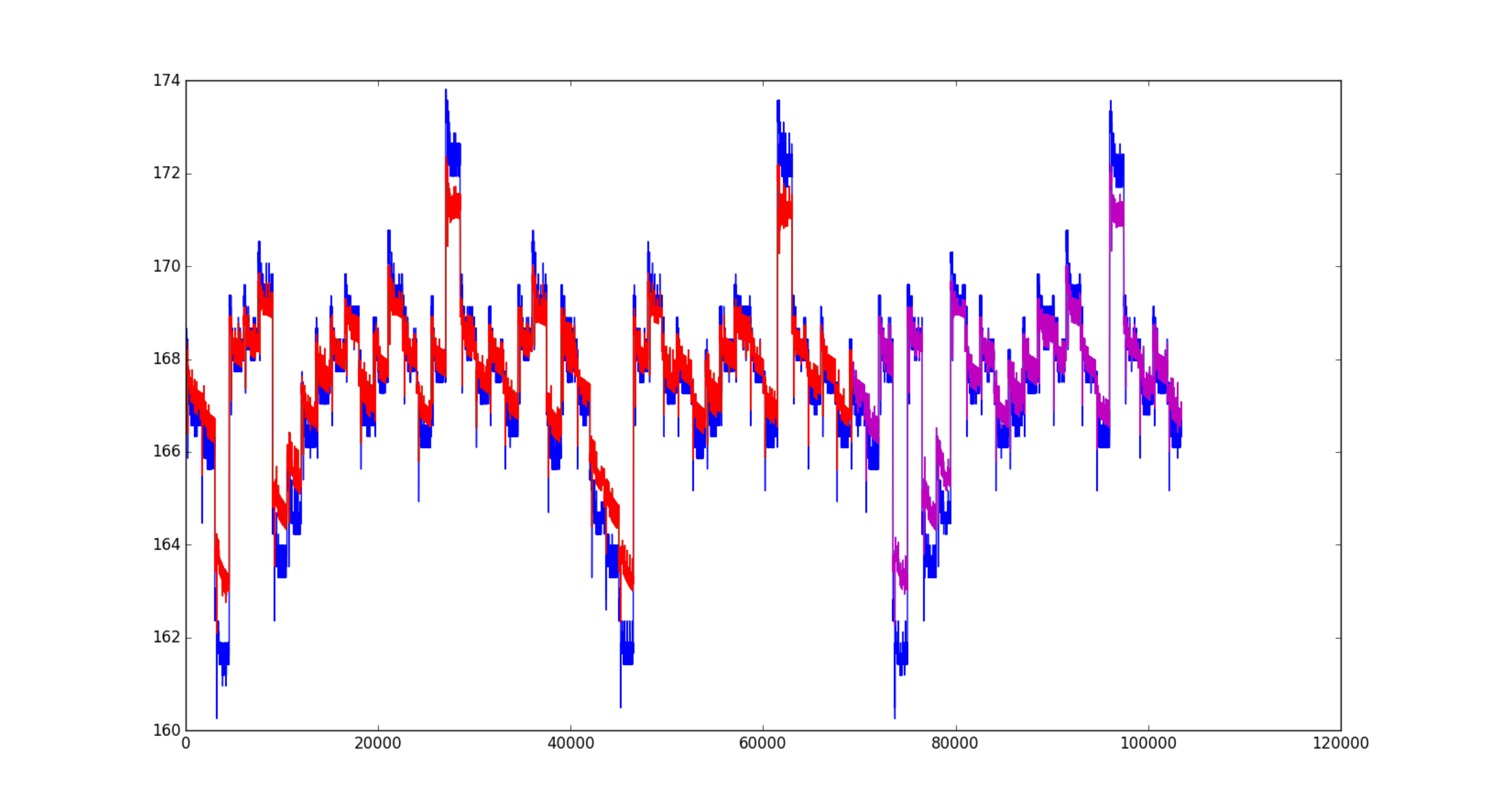}
		\caption{cells=16, epochs=2, RMSE = \SI{4.5}{\percent}}\label{fig:16_cells_2_epochs}
	\end{subfigure}
	\begin{subfigure}{\hsize}
	    \centering
		\includegraphics[width=0.8\hsize]{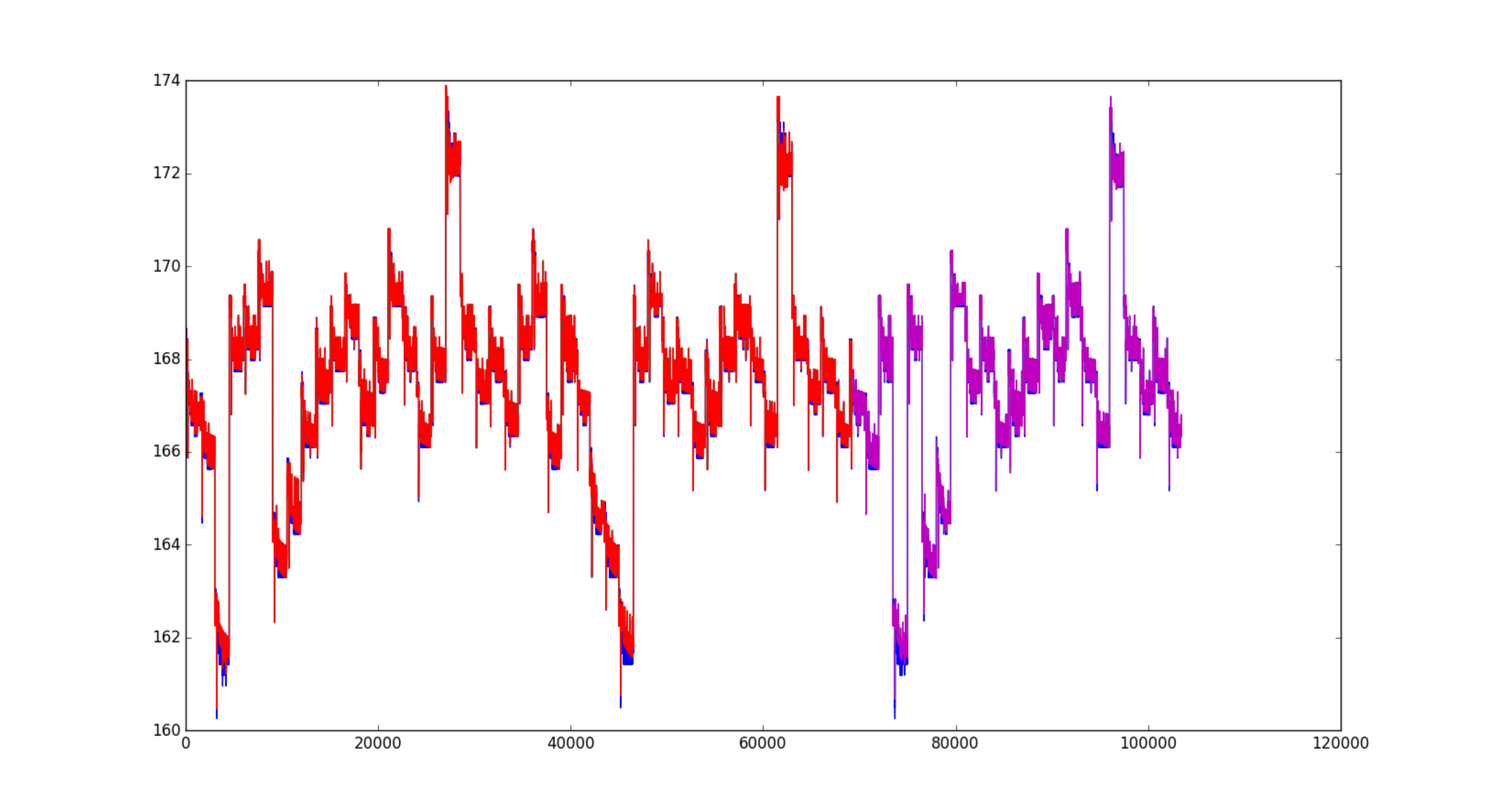}
		\caption{cells=16, epochs=6, RMSE = \SI{1.8}{\percent}}\label{fig:16_cells_6_epochs}
	\end{subfigure}
	\caption{Sample waveforms; single LSTM layer with look\_back=1, look\_ahead=1 and batch\_size=100; blue -- original signal, red -- train set prediction, magenta -- test set prediction}
\end{figure}

This section contains all the results of the experiments conducted to validate the feasibility of the application of the presented method. The learning process of the model consists of a series of steps, during which suitable parameters for obtaining the highest accuracy are selected. Fig.~\ref{fig:1_cell_2_epochs} -- \ref{fig:16_cells_6_epochs} present three examples of the results for different values of the hyper-parameters: number of cells and number of epochs.

Fig.~\ref{fig:1_cell_2_epochs} shows a virtually untrained \gls{LSTM} neural network and the results of its operation. The blue section highlights performance of the network on the training set, the red color denotes prediction results for the training set, and purple prediction results of the test set. \gls{RMSE} is almost \SI{16}{\percent}, which means a very large prediction error. 

Fig.~\ref{fig:16_cells_2_epochs} presents the results of the model for a network of \num{16} cells. Increasing the number of cells allowed for a much better results of \gls{RMSE} (\SI{4.5}{\percent}). It should be noted, however, that due to choosing only two epochs, the model did not manage to achieve their best performance and learn fully.

Increasing the amount of the epochs from two to six significantly improved the results of the model, which is reflected by the Fig.~\ref{fig:16_cells_6_epochs}. Consequently, \gls{RMSE} dropped to \SI{1.5}{\percent}, which means much better performance of the model. 

\begin{figure}
\includegraphics[width=\hsize]{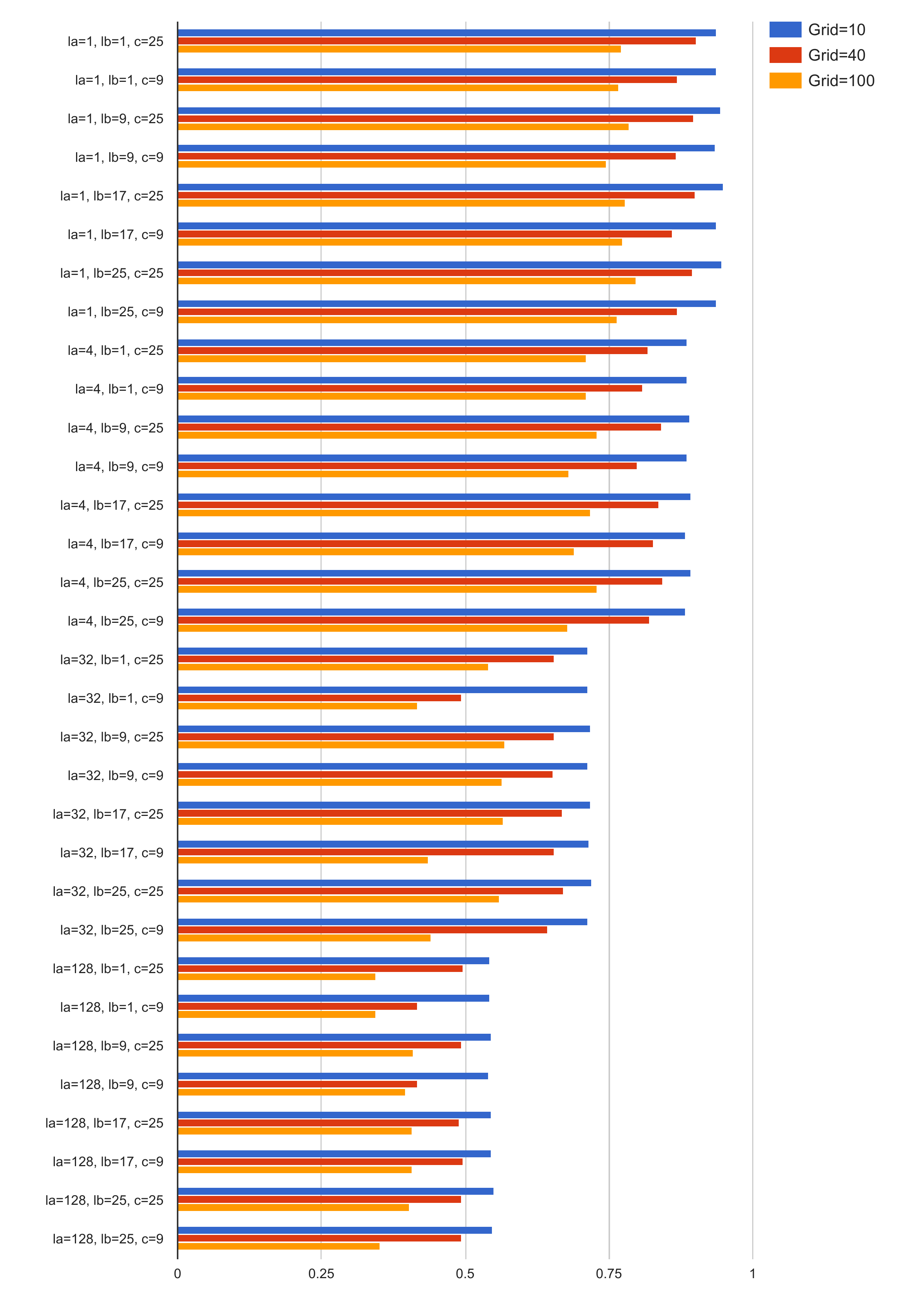}
\caption{Comparison of accuracy results depending on grid value for selected parameters: la -- look\_ahead, lb -- look\_back, c -- number of cells}\label{fig:pm_grid}
\end{figure}

\begin{figure}
\includegraphics[width=\hsize]{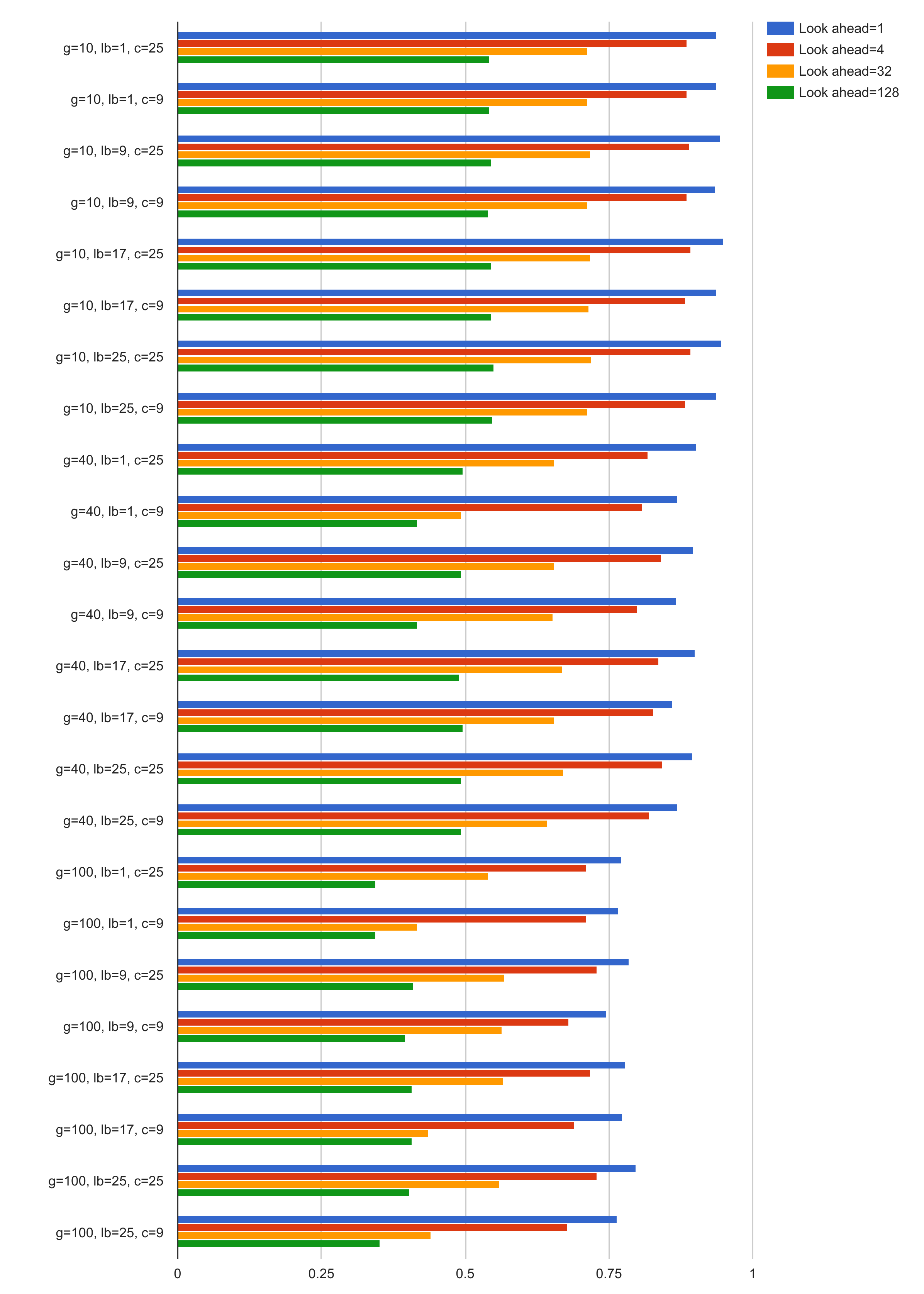}
\caption{Comparison of accuracy results depending on look ahead value for selected parameters: g -- grid size, lb -- look\_back, c -- number of cells}\label{fig:pm_la}
\end{figure}

\begin{figure}
\includegraphics[width=\hsize]{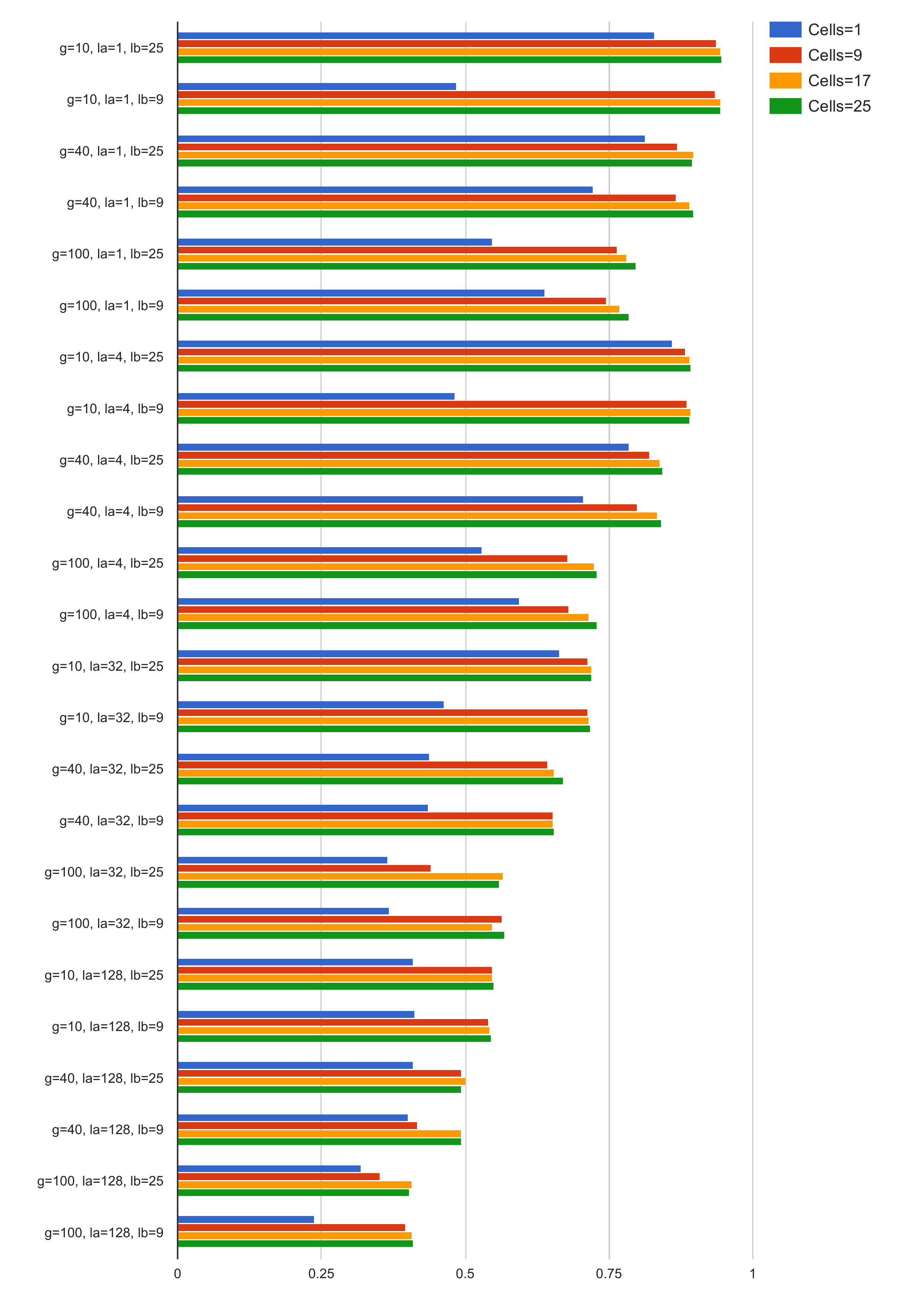}
\caption{Comparison of accuracy results depending on cells value for selected parameters: g -- grid size, la -- look\_ahead, lb -- look\_back}\label{fig:pm_cells}
\end{figure}

Next series of experiments was conducted for different values of grid size (g), look\_ahead steps (la), look\_back steps (lb) as well as the number of cells (c) in the \gls{LSTM} model. Batch size was fixed at \num{20}, with number of epochs being equal to \num{6}. The results of the experiments are presented in Tab.~\ref{tab:accuracy}, and Fig.~\ref{fig:pm_grid} -- \ref{fig:pm_cells}.

\begin{landscape}

\begin{table}
\caption{Prediction accuracy of $U_{\mathit{res}}$ signal using single layer LSTM; g -- grid size}
\label{tab:accuracy}
\centering
\begin{tabular}{cccccccccccccc}
\toprule
& &\multicolumn{3}{c}{look\_ahead=1}&\multicolumn{3}{c}{look\_ahead=4}&\multicolumn{3}{c}{look\_ahead=32}&\multicolumn{3}{c}{look\_ahead=128}\\
\cmidrule(lr){3-5} \cmidrule(lr){6-8} \cmidrule(lr){9-11} \cmidrule(lr){12-14}
look\_back&cells&g=10&g=40&g=100&g=10&g=40&g=100&g=10&g=40&g=100&g=10&g=40&g=100\\
\midrule
\multirow{4}{*}{$1$}&$1$&$0.925$&$0.789$&$0.386$&$0.876$&$0.745$&$0.383$&$0.667$&$0.437$&$0.367$&$0.411$&$0.386$&$0.318$\\
&$9$&$0.936$&$0.868$&$0.766$&$0.885$&$0.808$&$0.710$&$0.714$&$0.495$&$0.418$&$0.543$&$0.417$&$0.345$\\
&$17$&$0.939$&$0.903$&$0.767$&$0.885$&$0.817$&$0.710$&$0.714$&$0.655$&$0.540$&$0.543$&$0.495$&$0.346$\\
&$25$&$0.936$&$0.902$&$0.771$&$0.885$&$0.818$&$0.710$&$0.714$&$0.655$&$0.541$&$0.543$&$0.496$&$0.346$\\
\midrule
\multirow{4}{*}{$9$}&$1$&$0.484$&$0.722$&$0.638$&$0.483$&$0.706$&$0.594$&$0.464$&$0.437$&$0.368$&$0.412$&$0.401$&$0.238$\\
&$9$&$0.935$&$0.867$&$0.745$&$0.885$&$0.800$&$0.680$&$0.713$&$0.654$&$0.565$&$0.542$&$0.418$&$0.397$\\
&$17$&$0.944$&$0.890$&$0.769$&$0.892$&$0.835$&$0.715$&$0.716$&$0.652$&$0.548$&$0.543$&$0.494$&$0.407$\\
&$25$&$0.945$&$0.897$&$0.785$&$0.891$&$0.842$&$0.730$&$0.717$&$0.655$&$0.568$&$0.545$&$0.495$&$0.410$\\
\midrule
\multirow{4}{*}{$17$}&$1$&$0.853$&$0.717$&$0.542$&$0.483$&$0.691$&$0.530$&$0.464$&$0.433$&$0.368$&$0.479$&$0.402$&$0.319$\\
&$9$&$0.937$&$0.861$&$0.774$&$0.884$&$0.827$&$0.690$&$0.715$&$0.656$&$0.436$&$0.546$&$0.496$&$0.407$\\
&$17$&$0.944$&$0.899$&$0.758$&$0.892$&$0.828$&$0.694$&$0.714$&$0.654$&$0.557$&$0.546$&$0.496$&$0.407$\\
&$25$&$0.948$&$0.900$&$0.779$&$0.893$&$0.837$&$0.719$&$0.717$&$0.669$&$0.566$&$0.546$&$0.489$&$0.407$\\
\midrule
\multirow{4}{*}{$25$}&$1$&$0.829$&$0.812$&$0.549$&$0.861$&$0.786$&$0.530$&$0.664$&$0.437$&$0.367$&$0.411$&$0.410$&$0.318$\\
&$9$&$0.936$&$0.871$&$0.766$&$0.884$&$0.821$&$0.679$&$0.713$&$0.643$&$0.440$&$0.548$&$0.495$&$0.351$\\
&$17$&$0.944$&$0.897$&$0.782$&$0.891$&$0.838$&$0.726$&$0.720$&$0.655$&$0.566$&$0.547$&$0.500$&$0.408$\\
&$25$&$0.945$&$0.894$&$0.797$&$0.893$&$0.843$&$0.729$&$0.720$&$0.670$&$0.559$&$0.550$&$0.494$&$0.404$\\
\bottomrule
\end{tabular}
\end{table}

\end{landscape}

Fig.~\ref{fig:pm_grid} shows the values of accuracy for different grid sizes for various other parameters combinations. Analyzing the figure one can see that increasing the size of grid (reducing the single quantum size) leads to a deterioration of a model performance for the same parameters and the same set of data. This is the expected effect, which results from an increase in the number of categories that must be taken into consideration in the classification process while maintaining the existing network resources.

Fig.~\ref{fig:pm_la} focuses on the presentation of the results of the model depending on the value of the look\_ahead parameter. As expected, the more steps forward are anticipated, the lower accuracy is reached, because it is more challenging for the model to predict the correct categories. This effect even deepens with increase in the grid resolution and the network size reduction -- smaller net can not handle correct classification with not enough resources available. Since the look\_ahead parameter limits the anomaly detection window size, its value should be chosen carefully to allow for the best possible model accuracy while permitting a sufficiently large window size.

Fig.~\ref{fig:pm_cells} focuses on the presentation of the \gls{LSTM} model performance for a different number of cells. It is worth noting that without enough cells, and in particular using only one cell, the model is not able to accumulate all the training data dependencies needed to make the appropriate classification. It should also be noted that it is not necessary to use many more cells. In this case, using more than nine cells leads to very low improvement in the model performance. This observation leads to the conclusion that nine cells seem to be sufficient for this classification task.  

It should be emphasized that the proposed method introduces a clear way to determine whether a given set of model hyper-parameters is adequate for the task (achieves required accuracy for given predetermined grid and window sizes), while giving an opportunity to simplify the architecture as much as possible. This is critical due to the fact that the size of the network significantly affect the computational complexity of training and prediction. It is of great importance also in the case of hardware implementation of \gls{LSTM} and the \gls{GRU} networks, because of its size directly determines the amount of hardware resources to be used.

We also conducted experiment with the largest dataset of 302 MB and the following architecture of the network: single layer \gls{LSTM}, 128 cells, 20 epochs of training, look\_back=16, look\_ahead=128, grid=100 and optimizer \textit{Adam}. This resulted in a huge performance leap comparing to the results presented in Tab.~\ref{tab:accuracy}. The accuracy reached almost \SI{99.9}{\percent}. 

\begin{table}
\caption{Comparison of \gls{GRU} and \gls{LSTM} performance}
\label{tab:gru_lstm_comparison}
\centering
\begin{tabular}{lcc}
\toprule
&GRU&LSTM\\
\midrule
Cells&\num{16}&\num{16}\\
Epochs&\num{10}&\num{10}\\
Parameters&\num{864}&\num{1152}\\
Dataset (see Tab.~\ref{tab:datasets})&small&small\\
Accuracy [\si{\percent}]&\num{61.17}&\num{61.12}\\
Training time&\SI{2}{\hour}  \SI{13}{\minute}&\SI{2}{\hour} \SI{30}{\minute}\\
\bottomrule
\end{tabular}
\end{table}

We did most of our experiments using \gls{LSTM} algorithm for a sake of congruency and consistency with \cite{wielgosz2016usingLSTM}, which this paper is meant to be a continuation of in many aspects. Nevertheless, we decided do show the comparison between \gls{GRU} and \gls{LSTM} performance on a sample dataset as given in Tab.~\ref{tab:gru_lstm_comparison}.

\section{Conclusions and future work}
\label{section:conclusions}
This work extends existing experiments \cite{wielgosz2016usingLSTM} using higher resolution data and more diverse models.  As \gls{LHC} experiments enter \textit{High Luminosity} phase collision energies will be higher and more data will be collected what rises new challenges in maintenance of the equipment.

In experiments presented in this paper a $U_{\mathit{res}}$ signal was used. In the future experiments we plan on using several signals the same time and comparing performance with the one achieved in this paper. Nevertheless, a very promising results of \SI{99}{\percent} accuracy were achieved for the largest dataset of 302 MB and the following architecture of the network: single layer \gls{LSTM}, 128 cells, 20 epochs of training, look\_back=16, look\_ahead=128 and grid=100.

Another aspect worth investigating is feasibility of implementing predictive model on FPGAs. Performing computations on a PC works well for validation of the idea, but requirements of control systems like QPS are rather \textit{hard real-time} which PC systems are incapable of doing.

%\input{ack}

%\section*{References}
\bibliographystyle{elsarticle-num}
\bibliography{mybibfile,bibliography,authors-published}

\begin{thebibliography}{10}
\expandafter\ifx\csname url\endcsname\relax
  \def\url#1{\texttt{#1}}\fi
\expandafter\ifx\csname urlprefix\endcsname\relax\def\urlprefix{URL }\fi
\expandafter\ifx\csname href\endcsname\relax
  \def\href#1#2{#2} \def\path#1{#1}\fi

\bibitem{LHC_Nature}
O.~Br\"{u}ning, P.~Collier, {Building a behemoth}, NATURE 448 (2007)
  {285--289}.
\newblock \href {http://dx.doi.org/10.1038/nature06077}
  {\path{doi:10.1038/nature06077}}.

\bibitem{cern_crash}
\href{https://timeline.web.cern.ch/incident-at-the-lhc}{{Incident at the LHC}}
  [online] (2008) [cited 16-01-2017].

\bibitem{LHCDesRep}
L.~Evans, P.~Bryant, {LHC Machine}, Journal of Instrumentation 3~(08) (2008)
  S08001.
\newblock \href {http://dx.doi.org/10.1088/1748-0221/3/08/S08001}
  {\path{doi:10.1088/1748-0221/3/08/S08001}}.

\bibitem{SPSdesignReport}
\href{http://cds.cern.ch/record/110053}{{Report on the design study of a 300
  GeV proton synchrotron}}, Tech. rep., CERN (1964).
\newline\urlprefix\url{http://cds.cern.ch/record/110053}

\bibitem{ATLAS2008}
{ATLAS collaboration}, {The ATLAS Experiment at the CERN Large Hadron
  Collider}, JINST 3.

\bibitem{nieke2015analysis}
C.~Nieke, M.~Lassnig, L.~Menichetti, E.~Motesnitsalis, D.~Duellmann, {Analysis
  of CERN computing infrastructure and monitoring data}, Journal of Physics:
  Conference Series 664~(5) (2015) 052029.
\newblock \href {http://dx.doi.org/10.1088/1742-6596/664/5/052029}
  {\path{doi:10.1088/1742-6596/664/5/052029}}.

\bibitem{angelucci2014FPGA}
B.~Angelucci, R.~Fantechi, G.~Lamanna, E.~Pedreschi, R.~Piandani, J.~Pinzino,
  M.~Sozzi, F.~Spinella, S.~Venditti, {The FPGA based Trigger and Data
  Acquisition system for the CERN NA62 experiment}, Journal of Instrumentation
  9~(01) (2014) C01055.
\newblock \href {http://dx.doi.org/10.1088/1748-0221/9/01/C01055}
  {\path{doi:10.1088/1748-0221/9/01/C01055}}.

\bibitem{krizhevsky}
A.~Krizhevsky, I.~Sutskever, G.~E. Hinton,
  \href{http://papers.nips.cc/paper/4824-imagenet-classification-with-deep-convolutional-neural-networks.pdf}{{ImageNet
  Classification with Deep Convolutional Neural Networks}}, in: F.~Pereira,
  C.~J.~C. Burges, L.~Bottou, K.~Q. Weinberger (Eds.), {Advances in Neural
  Information Processing Systems 25}, Curran Associates, Inc., 2012, pp.
  1097--1105.
\newline\urlprefix\url{http://papers.nips.cc/paper/4824-imagenet-classification-with-deep-convolutional-neural-networks.pdf}

\bibitem{LeCun_deep_learning_2015}
Y.~LeCun, {Deep Learning of Convolutional Networks}, in: 2015 IEEE Hot Chips 27
  Symposium (HCS), 2015, pp. 1--95.
\newblock \href {http://dx.doi.org/10.1109/HOTCHIPS.2015.7477328}
  {\path{doi:10.1109/HOTCHIPS.2015.7477328}}.

\bibitem{graves2012supervised}
A.~Graves, Neural Networks, Springer Berlin Heidelberg, 2012.
\newblock \href {http://dx.doi.org/10.1007/978-3-642-24797-2}
  {\path{doi:10.1007/978-3-642-24797-2}}.

\bibitem{morton2016analysis}
J.~Morton, T.~A. Wheeler, M.~J. Kochenderfer, {Analysis of Recurrent Neural
  Networks for Probabilistic Modelling of Driver Behaviour}, IEEE Transactions
  on Intelligent Transportation Systems PP~(99) (2016) 1--10.
\newblock \href {http://dx.doi.org/10.1109/TITS.2016.2603007}
  {\path{doi:10.1109/TITS.2016.2603007}}.

\bibitem{pouladi2015recurrent}
F.~Pouladi, H.~Salehinejad, A.~M. Gilani, {Recurrent Neural Networks for
  Sequential Phenotype Prediction in Genomics}, in: 2015 International
  Conference on Developments of E-Systems Engineering (DeSE), 2015, pp.
  225--230.
\newblock \href {http://dx.doi.org/10.1109/DeSE.2015.52}
  {\path{doi:10.1109/DeSE.2015.52}}.

\bibitem{chen2016efficient}
X.~Chen, X.~Liu, Y.~Wang, M.~J.~F. Gales, P.~C. Woodland, {Efficient Training
  and Evaluation of Recurrent Neural Network Language Models for Automatic
  Speech Recognition}, IEEE/ACM Transactions on Audio, Speech, and Language
  Processing 24~(11) (2016) 2146--2157.
\newblock \href {http://dx.doi.org/10.1109/TASLP.2016.2598304}
  {\path{doi:10.1109/TASLP.2016.2598304}}.

\bibitem{zachary2015critical}
Z.~C. Lipton, J.~Berkowitz, C.~Elkan, {A Critical Review of Recurrent Neural
  Networks for Sequence Learning} (2015).
\newblock \href {http://arxiv.org/abs/1506.00019} {\path{arXiv:1506.00019}}.

\bibitem{greff2015lstm}
K.~Greff, R.~K. Srivastava, J.~Koutn{\'i}k, B.~R. Steunebrink, J.~Schmidhuber,
  {LSTM: A Search Space Odyssey} (2015).
\newblock \href {http://arxiv.org/abs/1503.04069} {\path{arXiv:1503.04069}}.

\bibitem{hochreiter1997long}
S.~Hochreiter, J.~Schmidhuber, {Long Short-Term Memory}, Neural Comput. 9~(8)
  (1997) 1735--1780.
\newblock \href {http://dx.doi.org/10.1162/neco.1997.9.8.1735}
  {\path{doi:10.1162/neco.1997.9.8.1735}}.

\bibitem{wielgosz2016usingLSTM}
M.~Wielgosz, A.~Skocze{\'{n}}, M.~Mertik, {Using LSTM recurrent neural networks
  for detecting anomalous behavior of LHC superconducting magnets} (2016).
\newblock \href {http://arxiv.org/abs/1611.06241} {\path{arXiv:1611.06241}}.

\bibitem{chung2015gated}
J.~Chung, C.~Gulcehre, K.~Cho, Y.~Bengio, {Gated Feedback Recurrent Neural
  Networks} (2015).
\newblock \href {http://arxiv.org/abs/1502.02367} {\path{arXiv:1502.02367}}.

\bibitem{chung2014empirical}
J.~Chung, C.~Gulcehre, K.~Cho, Y.~Bengio, {Empirical Evaluation of Gated
  Recurrent Neural Networks on Sequence Modeling} (2014).
\newblock \href {http://arxiv.org/abs/1412.3555} {\path{arXiv:1412.3555}}.

\bibitem{MPS_Wenninger}
J.~Wenninger, \href{https://arxiv.org/pdf/1608.03113.pdf}{{Machine Protection
  and Operation for LHC}}, CERN Yellow Report CERN-2016-002\href
  {http://arxiv.org/abs/arXiv:1608.03113v} {\path{arXiv:arXiv:1608.03113v}}.
\newline\urlprefix\url{https://arxiv.org/pdf/1608.03113.pdf}

\bibitem{interlocks}
F.~Bordry, R.~Denz, K.-H. Mess, B.~Puccio, F.~Rodriguez-Mateos, R.~Schmidt,
  \href{https://cds.cern.ch/record/531820/files/lhc-project-report-521.pdf}{{Machine
  Protection for the LHC: Architecture of the Beam and Powering Interlock
  System}}~({LHC Project Report 521}).
\newline\urlprefix\url{https://cds.cern.ch/record/531820/files/lhc-project-report-521.pdf}

\bibitem{MPS_Schmidt}
R.~Schmidt, \href{https://arxiv.org/pdf/1608.03087.pdf}{{Machine Protection and
  Interlock Systems for Circular Machines -- Example for LHC}}, CERN Yellow
  Report CERN-2016-002\href {http://arxiv.org/abs/arXiv:1608.03087v}
  {\path{arXiv:arXiv:1608.03087v}}.
\newline\urlprefix\url{https://arxiv.org/pdf/1608.03087.pdf}

\bibitem{Ciapala:691828}
E.~Ciapala, F.~Rodr\'{i}guez-Mateos, R.~Schmidt, J.~Wenninger,
  \href{http://cds.cern.ch/record/691828}{{The LHC Post-mortem System}}, Tech.
  Rep. LHC-PROJECT-NOTE-303, CERN, Geneva (Oct 2002).
\newline\urlprefix\url{http://cds.cern.ch/record/691828}

\bibitem{Lauckner:567214}
R.~J. Lauckner, \href{https://cds.cern.ch/record/567214}{{What data is needed
  to understand failures during LHC operation}}.
\newline\urlprefix\url{https://cds.cern.ch/record/567214}

\bibitem{Borland:1998}
M.~Borland, \href{http://www.aps.anl.gov/asd/oag/SDDSIntroTalk/slides.html}{{A
  Brief Introduction to the SDDS Toolkit}}, Tech. rep., Argonne National
  Laboratory, USA (1998).
\newline\urlprefix\url{http://www.aps.anl.gov/asd/oag/SDDSIntroTalk/slides.html}

\bibitem{CMW}
K.~Kostro, J.~Andersson, F.~{Di Maio}, S.~Jensen, N.~Trofimov, {The Controls
  Middleware (CMW) at CERN Status and Usage}, Proceedings of ICALEPCS,
  Gyeongju, Korea.

\bibitem{CMWnew}
A.~Dworak, F.~Ehm, P.~Charrue, W.~Sliwinski, {The new CERN Controls
  Middleware}, Journal of Physics Conference Series 396.

\bibitem{PMmigration}
{C.Aguilera-Padilla and S. Boychenko and M.Dragu and M.A. Galilee and J.C.
  Garnier and M. Koza and K. Krol and R. Orlandi and M.C. Poeschl and T.M.
  Ribeiro and M. Zerlauth}, {Smooth Migration of CERN POST MORTEM Service to a
  Horizontally Scalable Service}, Proceedings of ICALEPCS2015, Melbourne,
  Australia.

\bibitem{Andreassen:1235888}
O.~O. Andreassen, V.~Baggiolini, A.~Castaneda, R.~Gorbonosov, D.~Khasbulatov,
  H.~Reymond, A.~Rijllart, I.~Romera~Ramirez, N.~Trofimov, M.~Zerlauth,
  \href{https://cds.cern.ch/record/1235888}{{The LHC Post Mortem Analysis
  Framework}}, Tech. Rep. CERN-ATS-2010-009, CERN, Geneva (Jan 2010).
\newline\urlprefix\url{https://cds.cern.ch/record/1235888}

\bibitem{chollet2015}
F.~Chollet.
\newblock \href{https://github.com/fchollet/keras}{keras} [online] (2015).

\bibitem{articleELQA}
L.~Barnard, M.~Mertik, Usability of visualization libraries for web browsers
  for use in scientific analysis, International Journal of Computer
  Applications 121~(1) (2015) 1--5.
\newblock \href {http://dx.doi.org/10.5120/21501-4225}
  {\path{doi:10.5120/21501-4225}}.

\bibitem{mertikdhalerup}
M.~Mertik, K.~Dahlerup-Petersen, Data engineering for the electrical quality
  assurance of the lhc - a preliminary study, International Journal of Data
  Mining, Modelling and Management(in press).

\bibitem{Bokeh2016}
{Bokeh Development Team}.
\newblock \href{http://bokeh.pydata.org}{Bokeh: Python library for interactive
  visualization} [online] (2014) [cited 10.12.2016].

\bibitem{strecht2015comperative}
P.~Strecht, L.~Cruz, C.~Soares, J.~Mendes{-}Moreira, R.~Abreu,
  \href{http://www.educationaldatamining.org/EDM2015/proceedings/short392-395.pdf}{A
  comparative study of regression and classification algorithms for modelling
  students' academic performance}, in: Proceedings of the 8th International
  Conference on Educational Data Mining, {EDM} 2015, Madrid, Spain, June 26-29,
  2015, 2015, pp. 392--395.
\newline\urlprefix\url{http://www.educationaldatamining.org/EDM2015/proceedings/short392-395.pdf}

\end{thebibliography}

\end{document}